\documentclass[aip,graphicx]{revtex4-1}

\usepackage{geometry}
\usepackage{setspace}
\usepackage{fullpage}
\usepackage{color}
\usepackage{hyperref}
\usepackage[pdftex]{graphicx}
\draft 

\begin{document}

\title{Fe-based superconducting thin films on metallic substrates: growth, characteristics and relevant properties} 

\author{Kazumasa\,Iida}\email{iida@mp.pse.nagoya-u.ac.jp}
\affiliation{Department of Materials Physics, Nagoya University, Furo-cho, Chikusa-ku, Nagoya 464-8603, Japan}
\author{Jens\,H\"{a}nisch}
\affiliation{Institute for Technical Physics, Karlsruhe Institute of Technology, Hermann-von-Helmholtz-Platz 1, D-76344 Eggenstein-Leopoldshafen, Germany}
\author{Chiara\,Tarantini}
\affiliation{Applied Superconductivity Center, National High Field Laboratory, Florida State University, Tallahassee, FL, 32310, USA}

\date{\today}

\begin{abstract}
The discovery of Fe-based superconductors (FBS) as the second class of high-temperature superconducting transition (high--$T_{\rm c}$) materials after the cuprates generated a significant impact on the community of fundamental and applied superconductivity research. Whenever a new class of high--$T_{\rm c}$ superconducting materials is discovered, a lot of effort is devoted to growing single crystals and epitaxial thin films for exploring basic physical quantities. Although several properties of FBS are similar to the cuprates (large upper critical fields, as a consequence short coherence lengths, and small carrier density), others are distinctly different. For instance, in FBS the symmetry of the superconducting order parameter is most likely not a $d$--wave but an $s\pm$--wave, depending on the stoichiometry, crystallographic system, and doping level. Additionally, the critical current densities of FBS are less sensitive to the presence of grain boundaries (GBs) than those of the high--$T_{\rm c}$ cuprates. These features are highly beneficial for the realization of cheaper conductors for high-field magnets at low temperatures. Indeed, several groups have demonstrated FBS thin films on technical metallic substrates and powder-in-tube processed FBS wires as proof-of-principle studies for conductor applications. FBS on technical substrates also give many opportunities for studying how GB networks affect the critical current and how uniaxial strain impacts the superconducting properties. In this article, we review FBS thin films, especially on technical metallic substrates, and focus on application-relevant properties like pinning improvement by natural and artificial defects as well as the transparency of grain boundaries and GB networks. The recent development of FBS thin films on technical substrates and their superconducting properties  are presented and the performance gap with respect to films on single crystals is discussed.
\end{abstract}

\maketitle

\section{Introduction}
Since the first report of superconductivity in LaFeAs(O,F) 10 years ago by Hosono's group,\cite{Kamihara} a lot of progress in the growth and synthesis of Fe-based superconducting (FBS) single crystals, bulks, wires, and thin films has been made. Now, high-quality thin films [e.g., FeSe, Fe(Se,Te), doped and undoped $AE$Fe$_2$As$_2$, where $AE$ being an alkali earth element, and $Ln$FeAs(O,F), where $Ln$ is a lanthanoide element, named respectively {\it 11}, {\it 122} and {\it 1111}] are available on a large variety of single-crystalline and technical substrates. 
Hence fundamental as well as application-oriented research have been remarkably developed. In this article, we review FBS thin films, especially on technical metallic substrates, and focus on application-related properties like pinning improvement by natural and artificial defects as well as the transparency of grain boundaries (GBs) and GB networks. Many excellent review articles focussing on the growth of FBS thin films on single crystals and their physical properties have already been published\cite{Li, Hiramatsu-1, Mele, Haindl, Imai, Hosono-1, Hosono-2, Sakoda} and the interested readers may also refer to them.
The structure of this article is as follows: Naturally grown pinning centers are summarized, followed by artificial pinning centers. Afterwards, grain boundary experiments on Fe(Se,Te), doped BaFe$_2$As$_2$ (Ba-122) and NdFeAs(O,F) will be reviewed. Finally, FBS thin films on technical substrates will be described and discussed.

\section{Pinning improvements}
\begin{table}
\caption{Summary of $Gi$ for various FBS calculated by eq. (1) together with the physical quantities used for the calculations. The minimum flux creep rate $S$ at the possible operating temperature $T_{\rm op}$ is also shown. For comparison, the data for YBCO and MgB$_2$ are tabulated.}
 \begin{ruledtabular}
\begin{tabular}{ccccccc}
Materials                &  $T_{\rm c}$ (K)     & $\kappa$   & $\epsilon$   & $\mu_0H_{\rm c2}(0)$ (T) &  $Gi (\times10^{-3})$  & $S (\times10^{-2})$ at $T_{\rm op}$ \\  \hline
FeSe\cite{Yang}                                     &  8                 & 72               & 0.55       & 16      &  0.04                   &   0.3 at 4.2\,K            \\
Fe(Se$_{0.5}$Te$_{0.5}$)\cite{Klein}    &  14              & 287              & 0.27       & 116     &  17.2                  &   3.9 at 4.2\,K           \\
Co-doped Ba-122\cite{Yamamoto}        &  22               & 66               & 0.5         & 38       &  0.1                   &     0.2 at 4.2\,K         \\
P-doped Ba-122\cite{Chaparro}            &  29               & 93                & 0.4         &49       &  0.9                     &    0.4 at 4.2\,K          \\
K-doped Ba-122\cite{Yang}                   &  38               & 80                & 0.5         & 180     &  0.1                    &     0.6 at 20\,K         \\
NdFeAs(O,F)\cite{Kacmarcik}               &  35               & 113               & 0.13       & 60      &  20.6                   &     8.2 at 20\,K        \\
NdFeAs(O,F)\cite{Jaroszynski}             &  47               & 87                 & 0.11        & 162    &  6.5                     &    3.4 at 20\,K          \\
YBCO\cite{Yang}                                   &  91               & 62                 & 0.24        & 180    &  1.3                       &   3.0 at 77\,K           \\
MgB$_2$\cite{Zehetmayer}                   &  38               & 8                   & 0.23        & 3       &  $4\times10^{-3}$   &   0.1 at 20\,K           \\
\end{tabular}
\end{ruledtabular}
\label{tab:Gi}
\end{table}

One of the most important properties for wire and tape applications is the critical current density $J_{\rm c}$, which can be improved by microstructural engineering. Similar to the high--$T_{\rm c}$ cuprates, the coherence length of FBS is in the order of a few nm.\cite{Putti} Hence nano-sized precipitates and defects can strongly pin the vortices in the mixed state. When transport currents flow in FBS in the mixed state, the Lorentz force acts on the vortices. If the pinning force is weaker than the Lorentz force, the vortices start to move, resulting in dissipation of energy. Hence the stronger the pinning force density, the higher the critical current density. Another effect which depins vortices is thermal fluctuations. This is quantified by the Ginzburg number ($Gi$) defined as the ratio of the minimal condensation energy within the coherence volume and the thermal energy at the critical temperature ($k_{\rm B}T_{\rm c}$, where $k_{\rm B}$ is the Boltzmann constant). Thermal fluctuations are a serious problem for high--$T_{\rm c}$ superconductors (HTS), for instance in YBa$_2$Cu$_3$O$_7$ (YBCO) with a high $Gi$ of around 10$^{-3}$ (table\,\ref{tab:Gi}). For anisotropic superconductors, $Gi$ number in SI unit is evaluated by the following equation,\cite{Blatter}

\begin{equation}
Gi = \frac{1}{2}\left(\frac{k_{\rm B}T_{\rm c}}{4\pi\mu_0H_{\rm c}^2(0)\epsilon\xi^3(0)}\right)^2 = \frac{\pi\mu_0^2k_{\rm B}^2}{\phi_0^3}\frac{\kappa^4T_{\rm c}^2}{\mu_0H_{\rm c2}(0)\epsilon^2}= 1.07\times10^{-13}\frac{\kappa^4T_{\rm c}^2}{\mu_0H_{\rm c2}(0)\epsilon^2}
\end{equation}

\noindent
where $\mu_0H_{\rm c}(0)=\mu_0H_{\rm c2}(0)/\sqrt{2}\kappa$ is the thermodynamic critical field at 0\,K, $\mu_0H_{\rm c2}(0)$ is the upper critical field at 0\,K, $\epsilon$ is the anisotropy parameter defined as $\xi_{\rm c}(0)/\xi_{\rm ab}(0)$, $\xi_{\rm ab}(0)=\sqrt{\frac{\phi_0}{2\pi\mu_0H_{\rm c2}(0)}}$ is the in-plane coherence length at 0\,K, $\kappa$ is the Ginzburg-Landau parameter at 0\,K, and the right hand side is valid for basic SI units. It is noted that the $\mu_0H_{\rm c2}(0)$ is calculated by using the WHH formula. By using eq. (1), $Gi$ for various FBS is evaluated (table\,\ref{tab:Gi}). Only for {\it 122} and FeSe, $Gi$ is smaller than for YBCO and, therefore, the flux creep at the probable operating temperatures (minimum flux creep rate is expressed by $S\sim Gi^{1/2}(T/T_{\rm c}))$\cite{Eley} due to thermal fluctuations is not a serious problem. Different is the case of the most anisotropic NdFeAs(O,F) and Fe(Se,Te) that could more strongly suffer from thermal fluctuations.

\begin{figure}[ht]
	\centering
			\includegraphics[width=10cm]{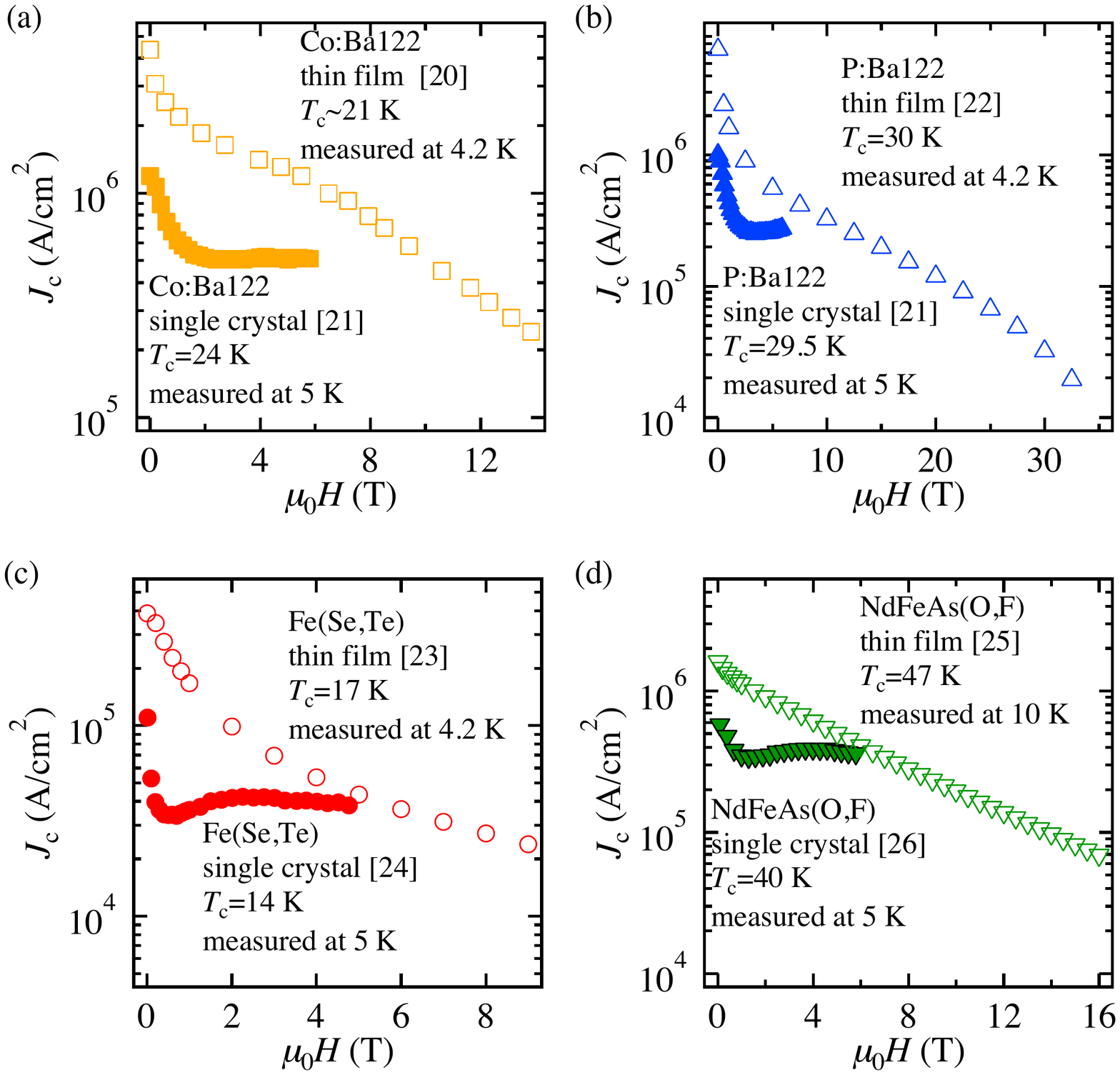}
		\caption{Comparison of $J_{\rm c}\mathchar`-H$ properties between FBS thin films and single crystals. The applied magnetic field was $H\parallel c$ and the measurement temperatures are shown in figure:
		(a) Co-doped Ba-122 on 50 unit cell BaTiO$_3$-buffered (LaAlO$_3$)$_{0.3}$(Sr$_2$TaAlO$_6$)$_{0.7}$ prepared by pulsed laser deposition (PLD)\,\cite{Lee-1} and optimally Co-doped Ba-122 single crystal.\cite{Ishida}
		(b) P-doped Ba-122 on MgO prepared by molecular beam epitaxy (MBE)\,\cite{Fritz} and optimally P-doped Ba-122 single crystal.\cite{Ishida}
		(c) Fe(Se,Te) on Fe-buffered MgO prepared by PLD\,\cite{Iida-0} and Fe$_{1+y}$Te$_{0.6}$Se$_{0.4}$ single crystal.\cite{Sun}
		(d) NdFeAs(O,F) on MgO fabricated by MBE\cite{Tarantini-6} and single crystal.\cite{Eisterer-0}} 
\label{fig:figure1}
\end{figure}

In general, thin films contain a higher density of natural defects than single crystals. In low magnetic fields, $J_{\rm c}$ for thin films is higher than those for single crystals (fig.\,\ref{fig:figure1}), whereas in high fields they appear to merge toward similar values. It is also noted that for single crystals $J_{\rm c}$ has a non-monotonic field dependence due to the fish tail effect, whereas such an effect is absent in thin films. Considering the good performance already obtained in clean thin films, improvements in the in-field behavior are achievable by the introduction of artificial pinning centers with an appropriate control of their dimensionality (1D: columnar defects and dislocations, 2D: grain boundaries and anti-phase boundaries, and 3D: particles\cite{Matsumoto}) and their density.

In this section, the naturally grown pinning centers and artificial pinning centers in FBS are reviewed, and how these pinning centers influence the in-field $J_{\rm c}$ is discussed. 

\subsection{Naturally grown pinning centers}     
\subsubsection{Ba-Fe-O nanopillars}

\begin{figure}[ht]
	\centering
			\includegraphics[width=7.5cm]{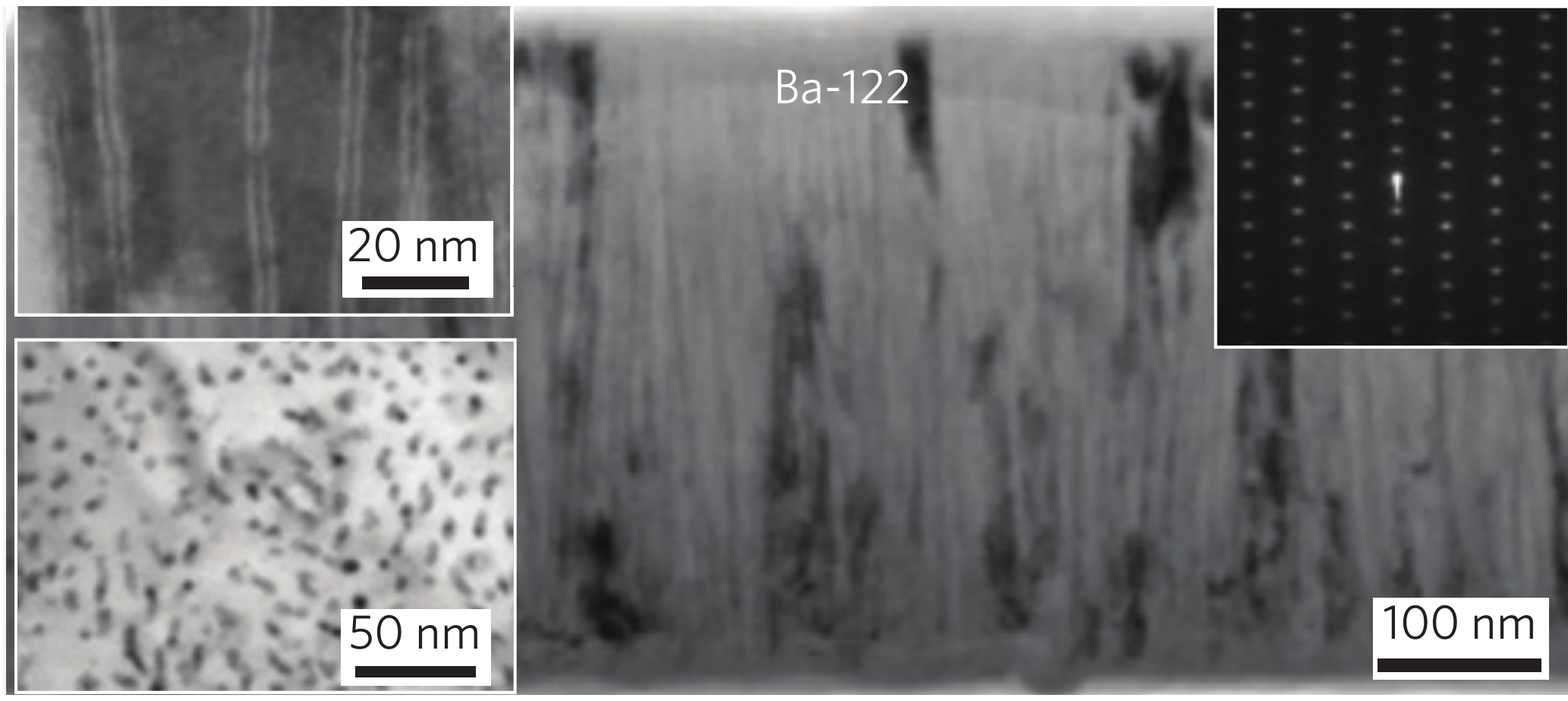}
		\caption{Cross-sectional and planer view of TEM images of a Co-doped Ba-122 thin film fabricated on SrTiO$_3$-buffered LSAT(001). Reproduced with permission from $Nat$. $Mater$. {\bf 9}, 397-402 (2010). Copyright 2010 Spring Nature.} 
\label{fig:figure2}
\end{figure}

S. Lee $et$ $al$. implemented SrTiO$_3$ as buffer layer to grow epitaxial Co-doped Ba-122 on 001-oriented substrates
((LaAlO$_3$)$_{0.3}$(Sr$_2$TaAlO$_6$)$_{0.7}$ (LSAT), GdScO$_3$, LaAlO$_3$ and Si) by pulsed laser deposition 
(PLD, KrF excimer laser).\cite{Lee-1} The nominal Co doping level of their PLD target was $x=0.08$ in Ba(Fe$_{1-x}$Co$_x$)$_2$As$_2$ with a small As excess to compensate the loss of As during the deposition. They concluded that the use of a SrTiO$_3$ template widened the range of possible substrates for high-quality Co-doped Ba-122 thin films. They also observed the formation of a high density of Ba-Fe-O nanopillars ($\sim8$\,vol.\%) that grew vertically from SrTiO$_3$ to the film surface (fig.\,\ref{fig:figure2}). As a result, the critical current density ($J_{\rm c}$) for $H\parallel c$ was higher than that for $H\parallel ab$, which differs from what is expected by the mass anisotropy. The mean distance between nanopillars was 16-17\,nm, which corresponds to a matching field of 7--8\,T.\cite{Tarantini-0} Indeed, the pinning force density showed a peak at around 8\,T for 
$H\parallel c$. Later it was found that the oxygen content of the PLD targets strongly affects the Ba-Fe-O nanopillar density.\cite{Tarantini-1} 
Most importantly, this compound can accept a high density of Ba-Fe-O of up to 20 vol.\% without compromising $T_{\rm c}$.

The detailed microstructural analysis by TEM identified the Ba-Fe-O phase as BaFeO$_2$, which is an isostructure of tetragonal SrFeO$_2$ ($a$=0.3991\,nm and $c$=0.3474\,nm).\cite{Zhang-Yi} The authors explained that BaFeO$_2$ has $a>0.3991$\,nm due to a larger ionic radius of Ba compared to Sr and is coherently strained by the Ba-122 matrix.

\subsubsection{Threading dislocations}

\begin{figure}[ht]
	\centering
			\includegraphics[width=12cm]{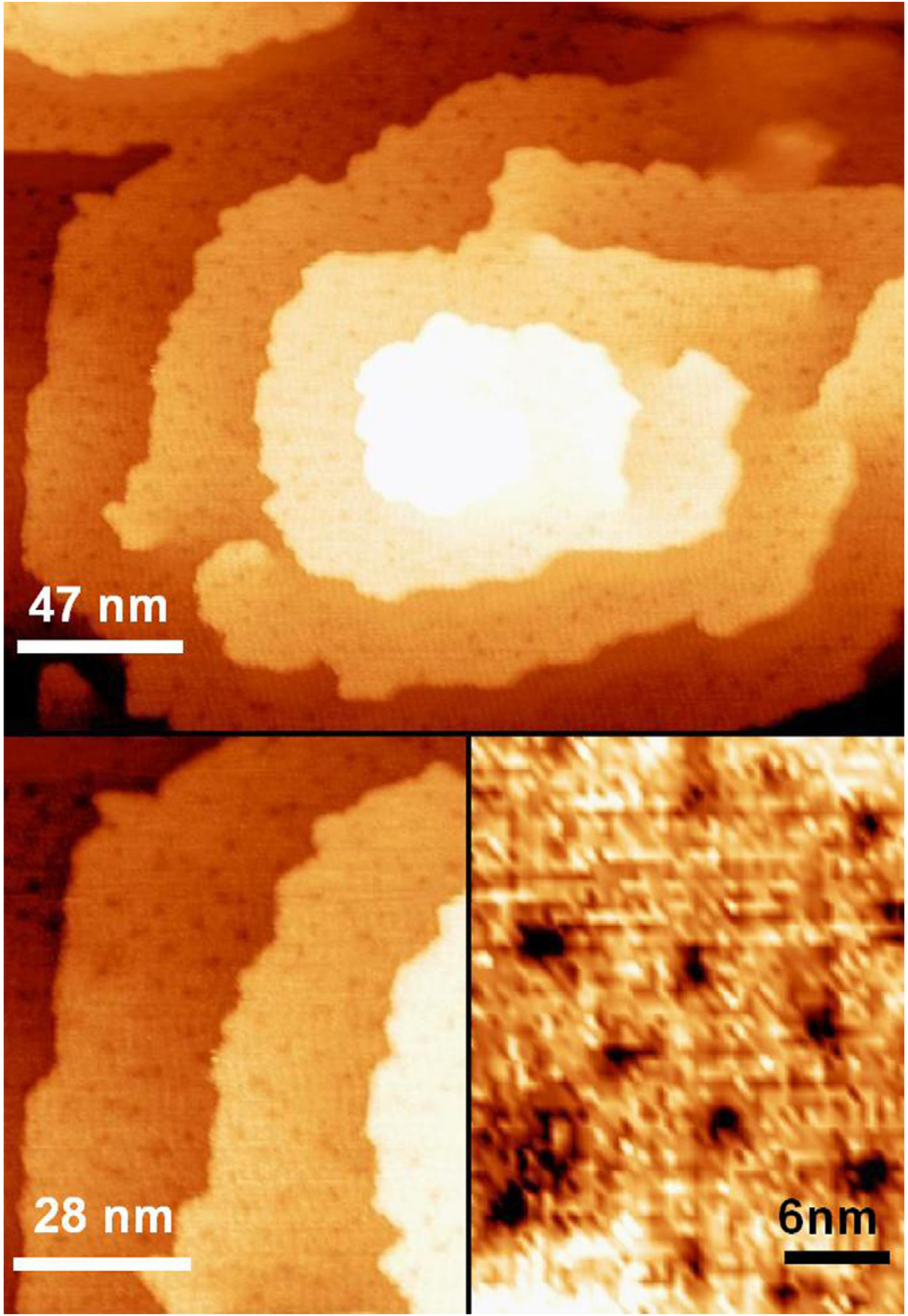}
		\caption{Surface morphology of a Fe(Se,Te) film deposited on SrTiO$_3$(001) substrate investigated by STM. Reproduced from $Appl$. $Phys$. $Lett$. {\bf100}, 082601 (2012), with the permission from AIP Publishing.} 
\label{fig:figure3}
\end{figure}

$c$-axis-correlated defects were also reported in Fe(Se,Te) thin films fabricated on SrTiO$_3$ (001) substrates by PLD (KrF excimer laser) in Genova. In fig.\,\ref{fig:figure3}, a high density of such defects with 2\,nm diameter, comparable to the coherence length, was observed.\cite{Bellingeri}
Bellingeri $et$ $al$. speculated those defects were threading dislocations. Owing to these defects, $J_{\rm c}$ for $H\parallel c$ was higher than for $H\parallel ab$ at all temperatures, which is similar to the observation reported for nanopillars in Co-doped Ba-122.\cite{Lee-1} The mean defect separation was about 10\,nm, corresponding to a matching field of 20\,T. Consequently, $J_{\rm c}$ for $H\parallel c$ has a weaker field dependence than that for $H\parallel ab$ up to the maximum measured field of 9\,T and in the temperature range below $0.65 T_{\rm c}$. The activation energy $U_0$ for vortex motion was raised by a factor 1.5 in these samples compared to a film on LaAlO$_3$(100) substrate, where interestingly this kind of defects was absent.

\begin{figure}[ht]
	\centering
			\includegraphics[width=7cm]{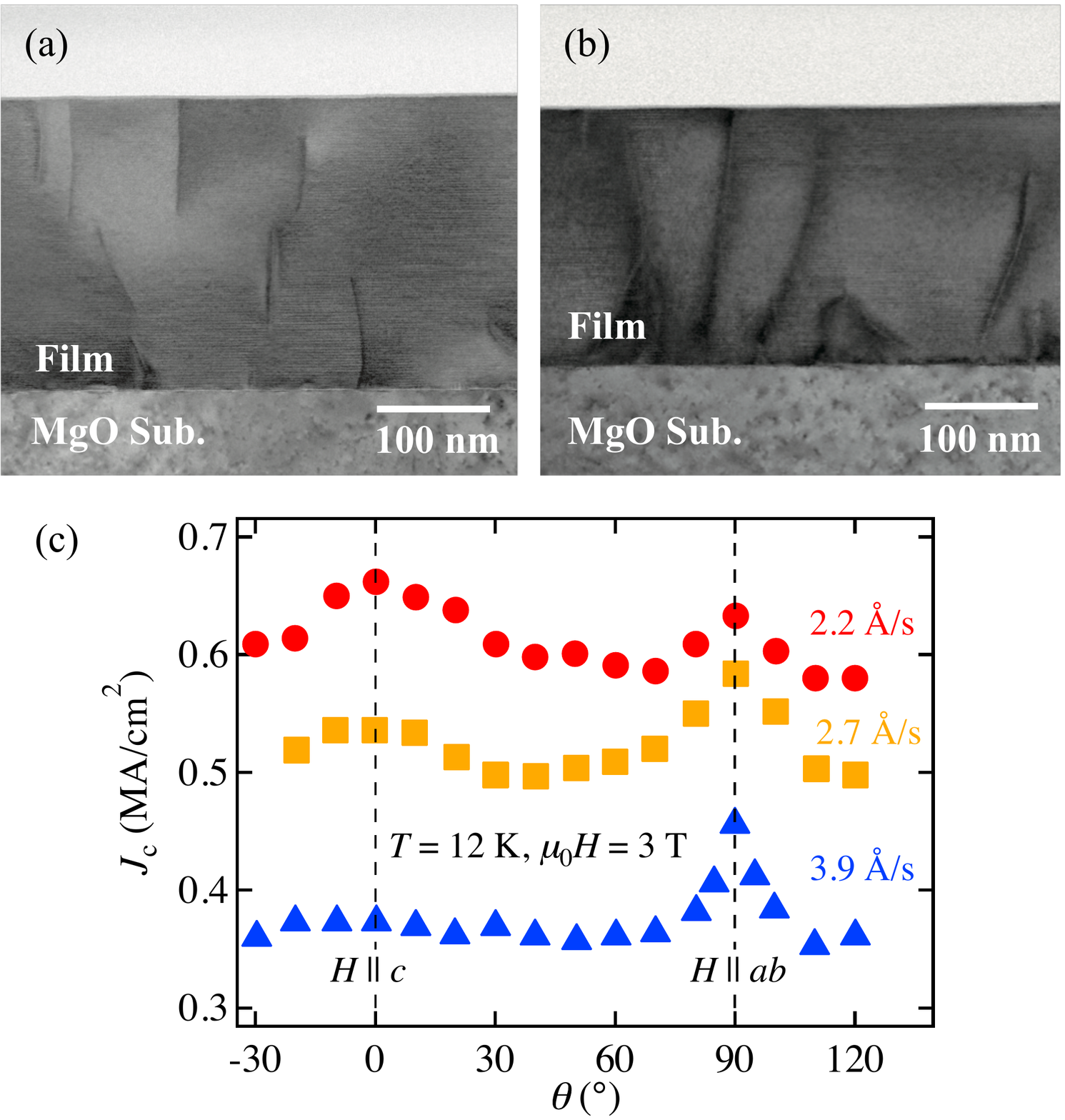}
		\caption{Microstructure and $J_{\rm c}$ anisotropy of P-doped Ba122 films with intentional threading dislocations. Cross-sectional view of P-doped Ba-122 thin films grown at a rate of (a) 2.2\,\AA/s and (b) 3.9\,\AA/s. (c) Angular dependence of $J_{\rm c}$ of P-doped Ba-122 deposited at various growth rate measured at 12\,K and 3\,T. The figure is redrawn with some modifications based on the data taken from fig.\,2 in Ref.\,\onlinecite{Sato-1}.} 
\label{fig:figure4}
\end{figure}

The introduction of threading dislocations in a controlled manner by tuning the growth conditions was reported for P-doped Ba-122 thin films by Sato $et$ $al$.\cite{Sato-1} Shown in fig.\,\ref{fig:figure4}(a) and (b) are the cross sectional views of two P-doped Ba-122 thin films grown on MgO substrates by PLD (Nd:YAG laser with 2nd harmonic) with different growth rates. Relatively short defects which are almost parallel to the Ba-122 $c$-axis were observed for a growth rate of 2.2\,\AA/s and identified as threading dislocations (fig.\,\ref{fig:figure4}(a)). On the other hand, tilted, long defects of relatively low density were seen for the film grown at 3.9\,\AA/s and identified as domain boundaries due to lateral growth (fig.\,\ref{fig:figure4}(b)). The film grown at 2.2\,\AA/s with threading dislocations showed the highest $J_{\rm c}$ in the full angular range among the films measured at 12\,K and 3\,T (fig.\,\ref{fig:figure4}(c)).

\subsubsection{Fe precipitates}     
\begin{figure}[ht]
	\centering
			\includegraphics[width=7cm]{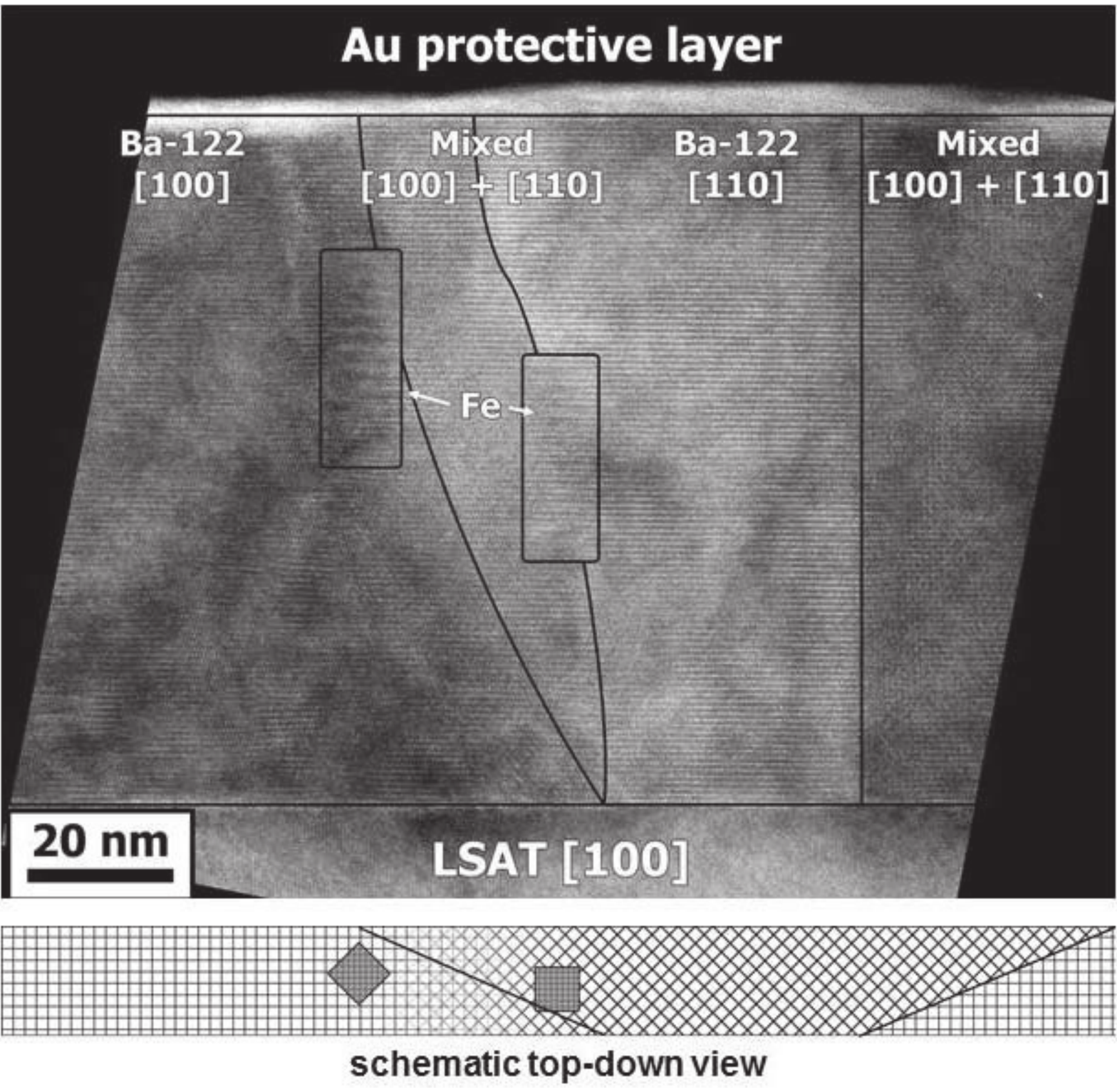}
		\caption{Cross sectional TEM image of the Co-doped Ba-122/LSAT film, revealing a sharp substrate-film interface, 45$^\circ$ [001] tilt GBs (vertical black lines) and round and 
		$c$-axis elongated Fe precipitates (encircled) near or within the GBs. The image was taken in the (100) pole. Below the TEM image is a schematic top-down view illustrating the orientation of GBs and Fe particles, deduced from Moir\'{e} contrasts. Reproduced from AIP Conference Proceedings {\bf1574}, 260 (2014), with the permission from AIP Publishing.} 
\label{fig:figure5}
\end{figure}

H\"{a}nisch $et$ $al$. investigated Co-doped Ba-122 thin films containing 45$^\circ$ rotated grains grown by PLD (KrF excimer laser).\cite{Jens} They fabricated Co-doped Ba-122 thin films on LSAT(001) at a relatively low deposition temperature of 675$^\circ$C in ultra-high vacuum (UHV) condition. As a result, round and $c$-axis elongated Fe precipitates were observed near or within the grain boundaries (fig.\,\ref{fig:figure5}). These Fe particles, together with the threading dislocations in the GBs lead to a $c$-axis peak in the $J_{\rm c}$ anisotropy.

For P-doped Ba-122 thin films grown by molecular beam epitaxy (MBE), Sakagami $et$ $al$. reported a maximum self-field $J_{\rm c}$ of 10\,MA/cm$^2$ (4.2\,K) for Fe-rich films (${\rm Fe}/{\rm Ba}=2.4$) on MgO(001) substrates measured by a magnetization method.\cite{Sakagami} Although no microstructural investigations have been reported, the authors speculated two plausible reasons: one possibility is that nanoparticles of Fe or Fe-based compounds, which were not detectable by x-ray diffraction (XRD), are dispersed in the matrix of such Fe-rich P-doped Ba-122 films and act as pining centers. The other possibility is that an Fe layer is formed at the interface with the substrate and acts
as a buffer layer that improves the crystallinity.

\subsection{Artificial pinning centers}  
\subsubsection{BaZrO$_3$ nanoparticles in PLD-grown Co- and P-doped Ba-122} 
Using the same strategy for pinning enhancement of YBa$_2$Cu$_3$O$_7$ films,\cite{Driscoll} BaZrO$_3$-added BaFe$_2$(As$_{1-x}$P$_x$)$_2$\,\cite{Miura} and Ba(Fe$_{1-x}$Co$_x$)$_2$As$_2$\,\cite{Lee-J} sintered pellets were prepared by two groups for use as PLD targets. The former target was ablated by a 2nd harmonic Nd:YAG laser whereas the latter was ablated by a KrF excimer laser.

\begin{figure}[ht]
	\centering
			\includegraphics[width=10cm]{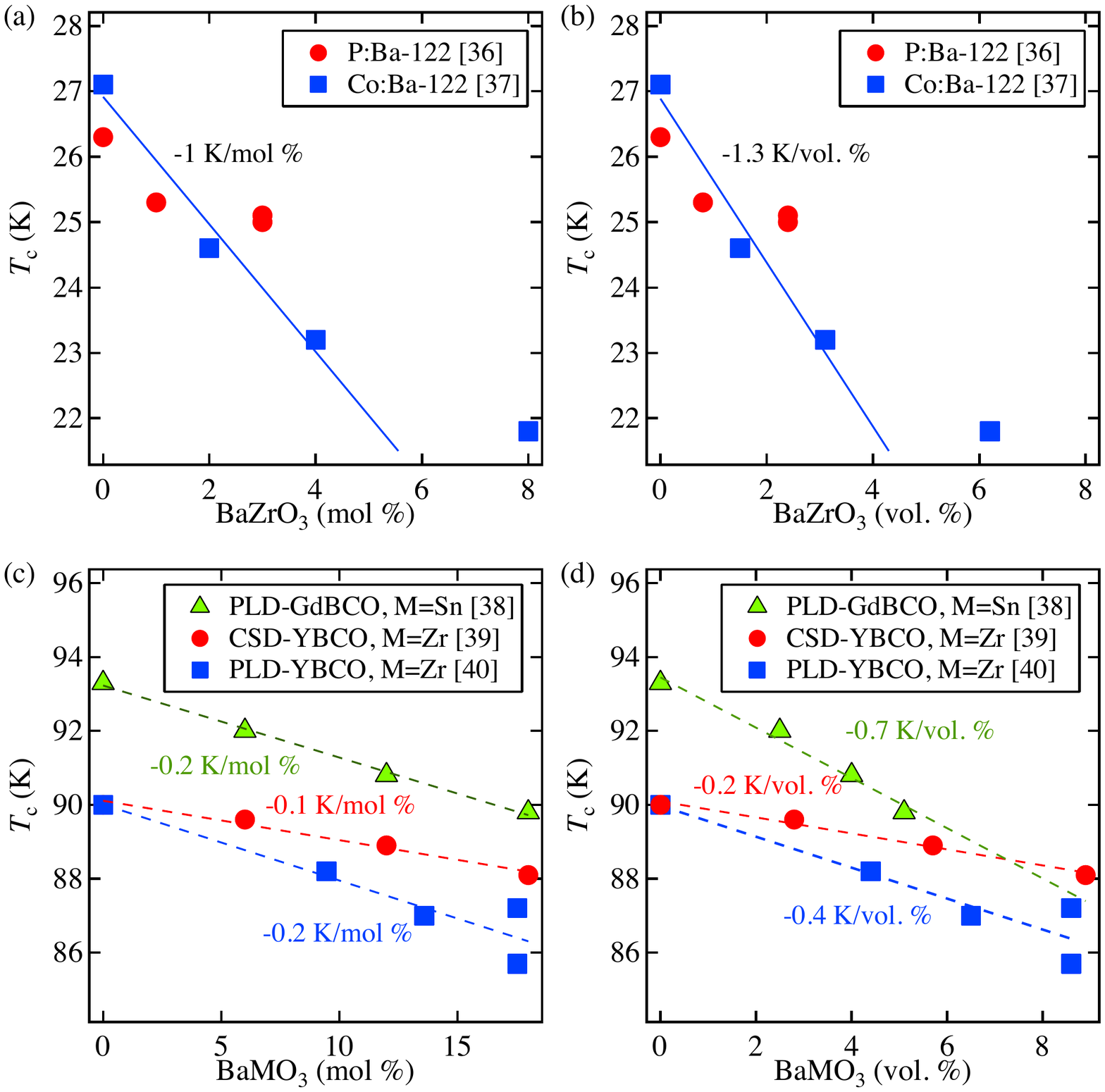}
		\caption{(a) Superconducting transition temperature of P-doped Ba-122 (P:Ba-122)\,\cite{Miura} and Co-doped Ba-122 (Co:Ba-122)\,\cite{Lee-J} as a function of BaZrO$_3$ content. For P:Ba-122, the zero resistance temperature $T_{\rm c,0}$ is plotted, whereas the onset $T_{\rm c,onset}$ is plotted for Co:Ba-122. (b) The same plot as a function of BaZrO$_3$ volume fraction. (c) Reduction of the $T_{\rm c}$ for GdBa$_2$Cu$_3$O$_7$ (GdBCO) grown by PLD\,\cite{Chepikov} and YBa$_2$Cu$_3$O$_7$ (YBCO) grown by chemical solution deposition (CSD)\,\cite{Birlik} and PLD\,\cite{Horide} as a function of BaMO$_3$ (M: Zr and Sn). (d) The same plot as a function of BaMO$_3$ volume fraction. Compared to the PLD-processed YBCO, the CSD-processed YBCO showed a gentle reduction of $T_{\rm c}$ around -0.1\,K/mol\% (-0.2\,K/vol.\%).} 
\label{fig:figure6}
\end{figure}

In P-doped Ba-122 films on MgO(001) substrate, nano-sized BaZrO$_3$ particles with average diameter of 5--10\,nm were finely distributed in the Ba-122 matrix.\cite{Miura} As a result, an almost isotropic $J_{\rm c}$ of 1.5\,MA/cm$^2$ at 15\,K ($0.6T_{\rm c}$) and 1\,T was obtained. In Co-doped Ba-122 films on CaF$_2$(001) substrate, short and $c$-axis-oriented BaZrO$_3$ nanorods with $\sim 4$\,nm diameter, slightly larger than in YBCO (2$\sim$ 3\,nm),\cite{Goyal} were formed in the Ba-122 matrix.\cite{Lee-J} The mean separation of BaZrO$_3$ nanorods was 10--11\,nm, corresponding to a matching field of 17--20\,T. The optimized Co-doped Ba-122 with 2\,mol\% BaZrO$_3$ showed an in-field $J_{\rm c}$ of 1.3\,MA/cm$^2$ at 13\,T and 4.2\,K.

Figure\,\ref{fig:figure6} compares the reduction of $T_{\rm c}$ in P-doped Ba-122 and Co-doped Ba-122. Up to 3 mol\% (2.5 vol.\%) of BaZrO$_3$, the initial suppression rate of $T_{\rm c}$ for both films is around -1\,K/mol\% (-1.3\,K/vol.\%), which is high compared to common values for GdBCO\,\cite{Chepikov} and YBCO\,\cite{Birlik, Horide}. The difference in $T_{\rm c}$ reduction against volume fraction is not as severe because of the smaller molar volume of the Ba-122 phase.

Compared to BaFeO$_2$ nanopillars (see section {\it Ba-Fe-O}), the reduction of $T_{\rm c}$ is significant, which may be ascribed to the different lattice mismatch.

\subsubsection{Superlattices and multilayers}
\begin{figure}[!hb]
	\centering
			\includegraphics[width=12cm]{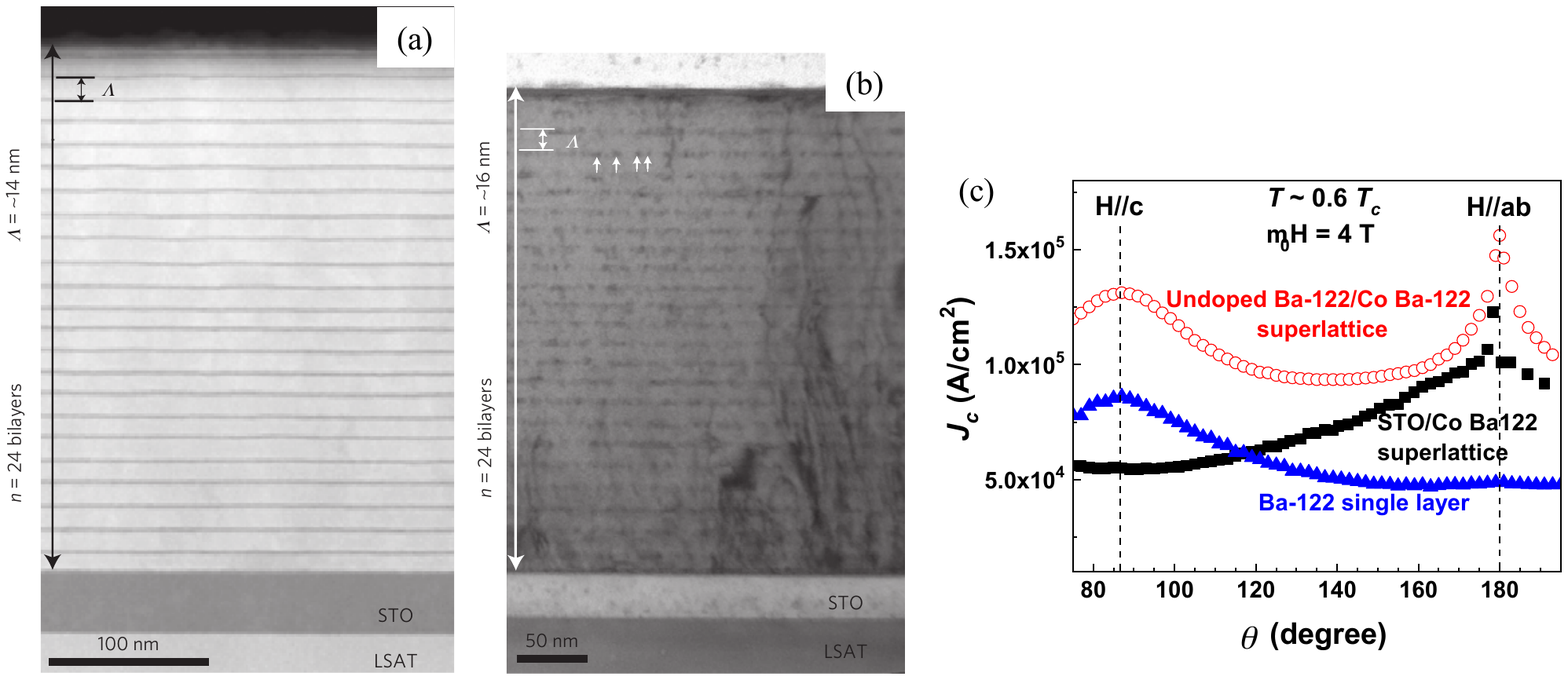}
		\caption{Cross sectional TEM image of (a) the SrTiO$_3$/Co-doped Ba-122 superlattice and (b) the undoped oxygen-rich Ba-122/Co-doped Ba-122 superlattice. Reproduced with permission from $Nat$. $Mater$. {\bf 12}, 392-396 (2013). Copyright 2013 Spring Nature.
		(c) Angular dependence of $J_{\rm c}$ of the SrTiO$_3$/Co-doped Ba-122 and  the undoped Ba-122/Co-doped Ba-122 superlattices measured at a reduced temperature of
		 $T$=0.6$T_{\rm c}$ and 4\,T. For comparison the data of the single layer of Co-doped Ba-122 was also plotted. The figure is redrawn with some modifications based on the data taken from fig.\,5c in Ref.\,\onlinecite{Lee-2}.} 
\label{fig:figure7}
\end{figure}

To date, FBS superlattices for Ba-122\,\cite{Lee-2} and Fe(Se,Te)\,\cite{Nabeshima} have been reported by the Univ. Wisconsin and the Univ. of Tokyo, respectively. S. Lee $at$ $al$. investigated the effect of SrTiO$_3$ and undoped Ba-122 insertion layers on the structural and superconducting properties of Co-doped Ba-122 thin films. The cross-sectional TEM images of the (100) projection clearly showed the 24 bilayers of SrTiO$_3$/Co-doped Ba-122 (fig.\,\ref{fig:figure7}(a)). On the other hand, the insertion layer of undoped Ba-122 grew discontinuously in the lateral direction (fig.\,\ref{fig:figure7}(b)). Similarly to the single layer reported by the same group, the undoped oxygen-rich Ba-122/Co-doped Ba-122 superlattice possessed vertical defects, presumably BaFeO$_2$, which act as effective pinning centers along the crystallographic $c$-direction. Combining the $c$-axis and $ab$-planar pinning contributions improved in-field $J_{\rm c}$ properties considerably, as shown in fig.\,\ref{fig:figure7}(c).

Nabeshima $et$ $al$. deposited FeSe/FeTe superlattices on CaF$_2$ and LaAlO$_3$.\cite{Nabeshima} From the microstructural investigations, some degree of Te/Se interdiffusion was observed. Because of the interdiffusion, the superconducting transition temperature of the FeSe/FeTe superlattice was improved compared to FeSe (i.e., from 12\,K for FeSe on CaF$_2$ to 15.8\,K for FeSe/FeTe superlattice on CaF$_2$).  No information on critical current densities was reported on these films.

Besides introducing artificial pinning centers, the multilayer approach can improve the superconducting properties.\cite{Tarantini-2,Engelmann-1, Chen-Li} Tarantini $et$ $al$. reported Co-doped, BaO-added Ba-122/Co-doped Ba-122 multilayer fabricated on CaF$_2$(001) substrate for which the superconducting transition temperature was increased by compressive strain.\cite{Tarantini-2} Pinning force densities as high as 84 and 70\,GN/m$^3$ for $H\parallel ab$ and $\parallel c$ at 4.2\,K were achieved by a combination of short $c$-axis nanorods, round nanoparticles and $ab$-arranged precipitates.

Engelmann $et$ $al$. fabricated
Co:Ba-122 (20\,nm)/Fe (4\,nm)/Co:Ba-122 (20\,nm)/Fe (30\,nm) multilayers on MgAl$_2$O$_4$, which showed a high $T_{\rm c}$ of 29.4\,K due probably to an As-deficiency.\cite{Engelmann-1} Chen $et$ $al$. reported the enhancement of in-field performance for an Fe(Se,Te) double layer with a CeO$_2$ insertion layer on SrTiO$_3$.\cite{Chen-Li} The improvement is ascribed to the additional interfacial defects introduced by the insertion layer.

\subsubsection{Irradiation}

\begin{table}[ht]
\caption{Radiation experiments on FBS thin films reported so far.}
\small
\begin{ruledtabular}
  \begin{tabular}{ccccc}
Materials                             &	     Particles  &	Total or Max. dose   &	 Change in $T_{\rm c}$ & Ref. \\ \hline
Fe(Se,Te) & Fast neutrons               &         $1.2\times 10^{17}\,{\rm cm}^{-2}$ &  -0.3\,K &  \onlinecite{Eisterer-2}\\
Fe(Se,Te)       & 190\,keV protons          &         $1.2\times 10^{17}\,{\rm cm}^{-2}$ & +0.5\,K  & \onlinecite{Ozaki}\\
Fe(Se,Te)       & 6\,MeV Au                 &         $1.0\times 10^{12}\,{\rm cm}^{-2}$ & -0.5\,K & \onlinecite{Ozaki-2}\\
Fe(Se,Te)       & 3.5\,MeV protons          &         $5.0\times 10^{15}\,{\rm cm}^{-2}$ & -7.0\,K & \onlinecite{Ahmad}\\
Fe(Se,Te)      & 3.5\,MeV protons          &         $7.3\times 10^{16}\,{\rm cm}^{-2}$ & -0.6\,K & \onlinecite{Sylva} \\
FeTe              & 150\,kV O                     &         $1.0\times 10^{17}\,{\rm cm}^{-2}$    & N.A. & \onlinecite{Janaki}\\
Co-doped Ba-122 & 3\,MeV protons       &         $2.0\times 10^{16}\,{\rm cm}^{-2}$ & -1.0\,K & \onlinecite{Boris}\\
Co-doped Ba-122  & 200\,keV protons &      $1.8\times 10^{15}\,{\rm cm}^{-2}$ & -18\,K & \onlinecite{Schilling}\\
P-doped Ba-122 & 250\,MeV Au &     $7.3\times 10^{11}\,{\rm cm}^{-2}$ &  -0.8\,K & \onlinecite{Daghero-1}\\
NdFeAs(O,F)      & 2\,MeV He &   $5.0\times 10^{15}\,{\rm cm}^{-2}$ & -3.0\,K & \onlinecite{Tarantini-4}\\
\end{tabular}
\end{ruledtabular}
\label{tab:radiation}
\end{table}
Although many irradiation experiments on FBS single crystals have been reported, only few studies except for {\it 11} report on the effects of irradiation in thin films (see table\,\ref{tab:radiation}). An excellent review article on radiation effects in FBS has recently been published by Eisterer.\cite{Eisterer-1} Here, we will briefly summarize the available data on FBS thin films.

The defects created by fast neutron irradiation on Fe(Se,Te) thin films produced only a small impact on the superconducting properties
 ($T_{\rm c}$, upper critical field, $H_{\rm c2}$, irreversibility field, $H_{\rm irr}$ and $J_{\rm c}$).\cite{Eisterer-2} On the other hand, low-energy proton irradiation created nanoscale compressive strain, resulting in an increase of $T_{\rm c}$ by 0.5\,K.\cite{Ozaki} Additionally, cascade defects combined with nanoscale strain worked as strong vortex pinning centers, and consequently $J_{\rm c}$ increased by one order of magnitude at 12\,K and by a factor of 2 at 4.2\,K at fields around and above 15\,T. Ozaki $et$ $al$. also investigated the effect of Au irradiation on the superconducting properties of Fe(Se,Te) thin films.\cite{Ozaki-2} The Au-irradiated film showed a slight decrease in $T_{\rm c}$ by 0.5\,K. However, in-field $J_{\rm c}$ properties were improved  for all field orientations at 4.2\,K. Furthermore, the $J_{\rm c}$ enhancement was nearly 70\% compared to the pristine film at 10\,K and 9\,T. Ahmad $et$ $al$. reported the effect of proton irradiation effect on the superconducting properties of FeSe$_{1-x}$Te$_x$ ($x$=0.4, 0.55) thin films.\cite{Ahmad} The reduction of superconducting properties caused by the local heating, which may accelerate the displacement of atoms and create vacancies. Sylva $et$ $al$. pointed out the importance of the position where the proton is implanted: the improvement of $J_{\rm c}$ up to 50\% at 12\,K and 7\,T was observed when protons were accumulated far from the film/substrate interface. Simultaneously, $T_{\rm c}$ and $H_{\rm c2}$ were unaltered.\cite{Sylva} On the other hand, superconducting properties were degraded when the proton implanted layer was close to the interface.

Non-superconducting FeTe thin films were irradiated by oxygen ions.\cite{Janaki} At the maximum dose of $1\times 10^{17}\,{\rm cm}^{-2}$, the FeTe film was amorphous.

The effect of 3\,MeV proton irradiation on the superconducting properties of Co-doped Ba-122 was reported by Maiorov $et$ $al$.\cite{Boris}
After the irradiation (total dose of $2\times10^{16}$\,cm$^{-2}$), the $T_{\rm c}$ decreased by only 1\,K. Overall $J_{\rm c}$ for the irradiated film was decreased in low field.
At high field and low temperatures, $J_{\rm c}$ was increased, particularly for $J_{\rm c}$ ($H\parallel ab$) due to a clear anisotropic contribution from the point-like defects.

The effect of proton irradiation on Co-doped Ba-122 thin films was also reported by Schilling $et$ $al.$ who investigated the disorder-induced symmetry change from $s\pm$ to $s++$ by THz spectroscopy.\cite{Schilling} On the assumption that the electron-electron coupling constant is positive, disorder induces a transition between $s\pm$ to $s++$ states.\cite{Efremov} In this report they observed such a transition: the small gap was closed at a certain amount of disorder and re-opened again by further increase of disorder. These results suggested that the superconducting order parameter in the pristine film has $s\pm$ symmetry. However, no information on $H_{\rm c2}$, $H_{\rm irr}$ and $J_{\rm c}$ was reported.

Daghero $et$ $al.$ reported on the effect of irradiation with 250-MeV Au ions on the superconducting properties of optimally P-doped Ba-122 thin films fabricated by MBE.\cite{Daghero-1} The residual resistivity at $T=0\,{\rm K}\,(\rho_0)$ increased with dose level, whilst the decrease in $T_{\rm c}$ was only 0.8\,K for a maximum dose of $7.26\times 10^{11}\,{\rm cm}^{-2}$. Such a small reduction in $T_{\rm c}$ would indicate $s\pm$ symmetry. The authors also considered the role of the substrate in the radiation effect: Au ion irradiation and heating in the substrate may have caused additional damage.

Unlike in NdFeAs(O,F) single crystals,\cite{Tarantini-3} $\alpha$-particle irradiation on NdFeAs(O,F) thin films did not create Kondo-like defects,\cite{Tarantini-4} suggesting that the type of defects induced by this type of irradiation is also strongly affected by other factors, the most likely being the initial composition. In the irradiated NdFeAs(O,F) films an increase of both $H_{\rm c2}$ and $H_{\rm irr}$ were observed. However, only for $H\parallel c$ a moderate improvement in $J_{\rm c}$  was found, whereas for $H\parallel ab$ point-defects compete with intrinsic pinning, reducing $J_{\rm c}$.

\section{Grain boundaries}
There are two main purposes for experiments on FBS thin films fabricated on bicrystal substrates having a single well-defined grain boundary.
One is to realize Josephson junctions with two FBS electrodes for fundamental research and assessing potential of electronics application. The other goal is to check the $J_{\rm c}$ transparency across the grain boundary, which is vital information for wire applications. In the former experiments, thin films are prepared on bicrystal substrates with relatively large misorientation angle $\theta_{\rm GB}$ to realize an SNS 
(superconductor--normal--superconductor) or SIS (superconductor--insulator--superconductor) junction. It is worth mentioning that all bicrystal experiments on FBS reported to date have been conducted using symmetric [001]-tilt bicrystal substrates (SrTiO$_3$, MgO and LSAT). Hence grain boundary experiments using different types of bicrystals (i.e., valley- or roof-type [010]-tilt and [100]-twist) are necessary to obtain deeper knowledge of grain boundary characteristics of FBS, since inter-grain $J_{\rm c}$ ($J_{\rm c}^{\rm GB}$) depends in general on the GB type, as shown for YBa$_2$Cu$_3$O$_7$
deposited on various types of bicrystals.\cite{Held} 
In general, different results are expected in case of different order parameter symmetry (e.g. $s$--wave or $d$--wave) as well as strain sensitivity and charge density.

To date, for FBS there have been only a few systematic and extended bicrystal experiments on Fe(Se,Te) and Co-doped Ba-122 films. Only singular experiments on P-doped Ba-122\,\cite{Sakagami} and recently on NdFeAs(O,F)\,\cite{Omura} have been reported so far, table\,\ref{tab:bi-crystal}. Reviews on GB properties in FBS bulk and film samples in comparison to
high-$T_{\rm c}$ cuprate superconductors have been given by Deutscher in 2010\,\cite{Deutscher} and Durrell $et$ $al.$ in 2011.\cite{John} In the following subsections, we review all reported bicrystal experiments on FBS concerning Josephson junctions followed by inter-grain $J_{\rm c}$ reduction due to weak-link behaviour. Some of the experiments mentioned below have been conducted for both purposes of Josephson junctions and investigation of the $J_{\rm c}$ transparency across the grain boundary. Therefore, occasionally the same experiments are referred to in the subsections {\it Josephson junctions with two FBS electrodes} and {\it Misorientation dependence of inter-grain $J_{\rm c}$}.

\begin{table}[htb]
\caption{Summary of bicrystal experiments on Fe-based superconductors. Note that all of these experiments have been conducted using [001]-tilt bicrystal substrates.}
\small
\begin{ruledtabular}
\begin{tabular}{ccccc}
Materials                &  Deposition  & Substrates & $\theta_{\rm GB}$ & $\theta_{\rm c}$ \\ \hline
Fe(Se,Te)               &  PLD   & SrTiO$_3$\,\cite{Sarnelli-1, Sarnelli-2, Sarnelli-3, Si-1} & $4^\circ \sim 45^\circ$ & 9$^\circ$\,\cite{Si-1,Sarnelli-2} \\
Co-doped Ba-122  & PLD    & SrTiO$_3$\,\cite{Lee-3, Schmidt-1} & $3^\circ \sim 30^\circ$ & \verb|<| 6$^\circ$\,\cite{Lee-3}\\
Co-doped Ba-122  & PLD    & MgO\,\cite{Katase-1, Hiramatsu-2, Iida-1}, LSAT\,\cite{Katase-1, Hiramatsu-2, Katase-junction-1, Katase-junction-2} & $3^\circ \sim 45^\circ$ & 9$^\circ$\,\cite{Katase-1, Hiramatsu-2}\\
P-doped Ba-122    &   MBE                           &   MgO\,\cite{Sakagami},  LSAT\,\cite{Schmidt-2} & $24^\circ$,  $45^\circ$ & N.A.\\
NdFeAs(O,F)         &  MBE                             &   MgO\,\cite{Omura} & $6^\circ \sim 45^\circ$ & \verb|<| 6$^\circ$ \\
\end{tabular}
\end{ruledtabular}
\label{tab:bi-crystal}
\end{table}

\subsection{Josephson junctions with two FBS electrodes}
\subsubsection{Fe(Se,Te)}
To date, four bicrystal experiments on Fe(Se,Te) have been published. First bicrystal experiments on Fe(Se,Te) were reported by Sarnelli $et$ $al.$ who fabricated 150\,nm thick Fe(Se,Te) films on [001]-tilt SrTiO$_3$ bicrystals with a misorientation angle ($\theta_{\rm GB}$) of 45$^\circ$ by PLD using a Nd:YAG laser with a wave length of 1024\,nm.\cite{Sarnelli-1} The authors made microbridges (2.6--20\,$\mu$m wide and 20\,$\mu$m long) on Fe(Se,Te) bicrystals by optical lithography and ion beam milling. To avoid possible damage during the process, the sample holder was cooled to -40$^\circ$C. The $I\mathchar`-V$ characteristics at 4.2\,K showed a resistively shunted junction (RSJ) behavior with a low normal-state resistance $R_{\rm N}$ in the range $40\leq R_{\rm N}\leq 270\,{\rm m}\Omega$, which is typical for SNS junctions. Compared to the cuprates, those values are two orders of magnitude lower. As a result, the $I_{\rm c}R_{\rm N}$ product was 18$\sim$33\,$\mu$V, where $I_{\rm c}$ is the Josephson critical current. Later, Sarnelli $et$ $al.$ fabricated dc superconducting quantum interference devices (dc-SQUIDs) using 100-nm thick Fe(Se,Te) bicrystal with $\theta_{\rm GB}=24^\circ$.\cite{Sarnelli-2, Sarnelli-3} The resultant device showed a clear voltage modulation under magnetic field with a maximum amplitude of 3.8\,$\mu$V at 4.2\,K.

Si $et$ $al$. fabricated 100-nm thick Fe(Se,Te) thin films on CeO$_2$-buffered SrTiO$_3$ bicrystal substrates ($\theta_{\rm GB}$=4$^\circ$, 7$^\circ$, 15$^\circ$ and 24$^\circ$) by PLD.\cite{Si-1} Microbridges with 300\,$\mu$m length and 20--25\,$\mu$m width were fabricated by laser cutting for transport measurements. The $\rho\mathchar`-T$ curve of the bridge with $\theta_{\rm GB}=24^\circ$ showed a double superconducting transition at 20\,K and 17.5\,K with $\rho$=0 at around 12\,K. On the other hand, the bridge with $\theta_{\rm GB}=4^\circ$ had one transition at 20\,K similar to the intra-grain bridge. Using the bridge with $\theta_{\rm GB}=24^\circ$, the authors observed  multiple Josephson current modulations by magnetic field. The magnitude of the $I_{\rm c}$ modulation was around 80\%.
The $I\mathchar`-V$ characteristics at 2\,K roughly followed the RSJ model with a normal state resistance of $R_{\rm N}$=28\,m$\Omega$.
Such a low value of $R_{\rm N}$ is due to the metallic nature of the normal state of the Fe(Se,Te) bicrystal grain boundary junction. The $I_{\rm c}R_{\rm N}$ product of their junctions was estimated to around 22\,$\mu$V at 2\,K.

\subsubsection{Co- and P-doped Ba-122}
Katase $et$ $al.$ reported on Josephson junctions and dc-SQUIDs in PLD-processed Co-doped Ba-122 ($T_{\rm c}$=21.5--22.6\,K) using LSAT bicrystal substrates with $\theta_{\rm GB}=30^\circ$.\cite{Katase-junction-1, Katase-junction-2} For single Josephson junctions, the authors fabricated microbridges (10\,$\mu$m wide and 300\,$\mu$m long) by photolithography and Ar ion milling. The $I\mathchar`-V$ characteristics of the resultant junction at 4.2\,K showed RSJ behavior. The normal state resistance $R_{\rm N}$ of the junction was around 12\,m$\Omega$, resulting in a low $I_{\rm c}R_{\rm N}$ product of $\sim$50\,$\mu$V. For dc-SQUIDs, bridges with a smaller width of 3\,$\mu$m (area of SQUID loop was 18$\times$8\,$\mu$m$^2$) were made to reduce the critical current. The resultant device showed a clear voltage modulation with an amplitude of 1.4\,$\mu$V at 14\,K. However, the shape of $V\mathchar`-\Phi$ deviated from the ideal sinusoidal function. Additionally, the level of flux noise at 1\,Hz was $\sim4.3\times10^{-4}$\,$\Phi_0$Hz$^{-1/2}$, which was one order of magnitude higher than for cuprates. From those junction experiments, 
the barrier layer is of metallic nature in Co-doped Ba-122. Indeed, the inter-grain $J_{\rm c}^{\rm GB}$ showed a  quadratic temperature dependence, indicative of SNS rather than SIS junctions.\cite{Katase-1} Hence, for realizing practical SQUIDs it is necessary to increase the junction resistance $R_{\rm N}$ by an artificial insulating barrier.

Schmidt $et$ $al$. also reported on PLD-grown Co-doped Ba-122 grain boundary junctions using SrTiO$_3$ with $\theta_{\rm GB}=30^\circ$.\cite{Schmidt-1} To suppress oxygen diffusion from SrTiO$_3$, a 10\,nm thick MgAl$_2$O$_4$ layer was deposited on the SrTiO$_3$ bicrystal. Fe-buffer layer followed by Co-doped Ba-122 were deposited on this template. All layers were epitaxially grown as confirmed by XRD. In their junctions ($T_{\rm c}$=24\,K), a high excess current ($I_{\rm ex}=645$\,$\mu$A) was observed, and therefore a small $I_{\rm c}R_{\rm N}$ product of 6.5\,$\mu$V at 4.2\,K was obtained.

Sakagami $et$ $al$. prepared P-doped Ba-122 by MBE on [001]-tilt MgO bicrystals with $\theta_{\rm GB}=24^\circ$.\cite{Sakagami}
The $I\mathchar`-V$ characteristics of their device ($T_{\rm c}$=29.5\,K, bridge width of 30\,$\mu$m) at 2\,K was described by a flux-flow extension of the RSJ model, which gave a normal state resistance of $R_{\rm N}$=4.4\,m$\Omega$ and a Josephson critical current of $I_{\rm c}$=8.5\,mA. As a consequence, a small $I_{\rm c}R_{\rm N}$ product of 37\,$\mu$V was obtained. Such a low normal state resistance of the junction is due to the metallic nature of the grain boundary. The same group also fabricated P-doped Ba-122 thin films on LSAT bicrystals with $\theta_{\rm GB}=45^\circ$\,\cite{Schmidt-2}. The $I\mathchar`-V$ characteristics of the resultant junction can again be fitted by a RSJ model with flux flow current. Their device had an $I_{\rm c}R_{\rm N}$ product of 45\,$\mu$V at 4.2\,K. However, this value was corrected to 11\,$\mu$V by taking into account the excess current.

\subsubsection{Summary of Josephson junctions with two FBS electrodes}
All bicrystal junctions of Fe(Se,Te), Co-doped, and P-doped Ba-122 have the same low level of $I_{\rm c}R_{\rm N}$ (tens of $\mu$V) compared to that of YBCO (in the order of mV at 4.2\,K). The main reason for such low $I_{\rm c}R_{\rm N}$ is that the barrier layer is of metallic nature, which leads to a low junction resistance $R_{\rm N}$. Therefore, the dc-SQUIDs made from Fe(Se,Te)\,\cite{Sarnelli-2} and Co-doped Ba-122\,\cite{Katase-junction-2} showed high flux noise. The properties mentioned above are not suitable for electronics applications, however, are highly beneficial to wire applications. To realize electronics applications like SQUIDs, one would have to increase the $R_{\rm N}$ by an artificial insulating barrier. Interested readers for Josephson effects in FBS may refer to Ref.\,\onlinecite{Seidel-1}.

\subsection{Misorientation dependence of inter-grain $J_{\rm c}$}
\subsubsection{Fe(Se,Te)}
 
The inter-grain $J_{\rm c}$ of Fe(Se,Te) across a $\theta_{\rm GB}$=45$^\circ$ grain boundary without magnetic field was around $\sim10^4$\,A/cm$^2$ at 4.2\,K, one order of magnitude lower than the intra-grain $J_{\rm c}$ ($J_{\rm c}^{\rm GB}$).\cite{Sarnelli-1} These results suggested that the decay of $J_{\rm c}$ with GB angle may not be as strong as for the cuprates, where at this angle the suppression is significant (4 orders of magnitude). Later, Sarnelli $et$ $al$. have reported $J_{\rm c}^{\rm GB}/J_{\rm c}^{\rm Grain}$ as a function of $\theta_{\rm GB}$.\cite{Sarnelli-2} It was found that the $\theta_{\rm GB}$ dependence of $J_{\rm c}^{\rm GB}/J_{\rm c}^{\rm Grain}$ is almost similar to the result reported by Si $et$ $al$.\cite{Si-1} Additionally, the level of $J_{\rm c}^{\rm GB}$ was almost constant for $\theta_{\rm GB}=24^\circ$ and $\theta_{\rm GB}=45^\circ$, which will be discussed later.

Si $et$ $al.$ found that $J_{\rm c}^{\rm GB}$ for $\theta_{\rm GB}=4^\circ$ and 7$^\circ$ maintained $\sim10^5$\,A/cm$^2$ even in the presence of large 
applied magnetic fields ($\mu_0H=10\,{\rm T}$, $H\parallel c$) at 4.2\,K, indicative of strong links in this $\theta_{\rm GB}$ range.\cite{Si-1} On the other hand, $J_{\rm c}^{\rm GB}$ for $\theta_{\rm GB}>15^\circ$ was significantly suppressed by a small magnetic field: $J_{\rm c}^{\rm GB}$ for $\theta_{\rm GB}=24^\circ$, e.g., was reduced from $4\times10^4$ to $7\times10^3$\,A/cm$^2$ even by a small magnetic field of $\mu_0H=0.5\,{\rm mT}$. Based on the experimental results above, the authors concluded that the critical angle $\theta_{\rm c}$ was around 9$^\circ$.

\subsubsection{Co- and P-doped Ba-122}
\begin{figure}[ht]
	\centering
			\includegraphics[width=8cm]{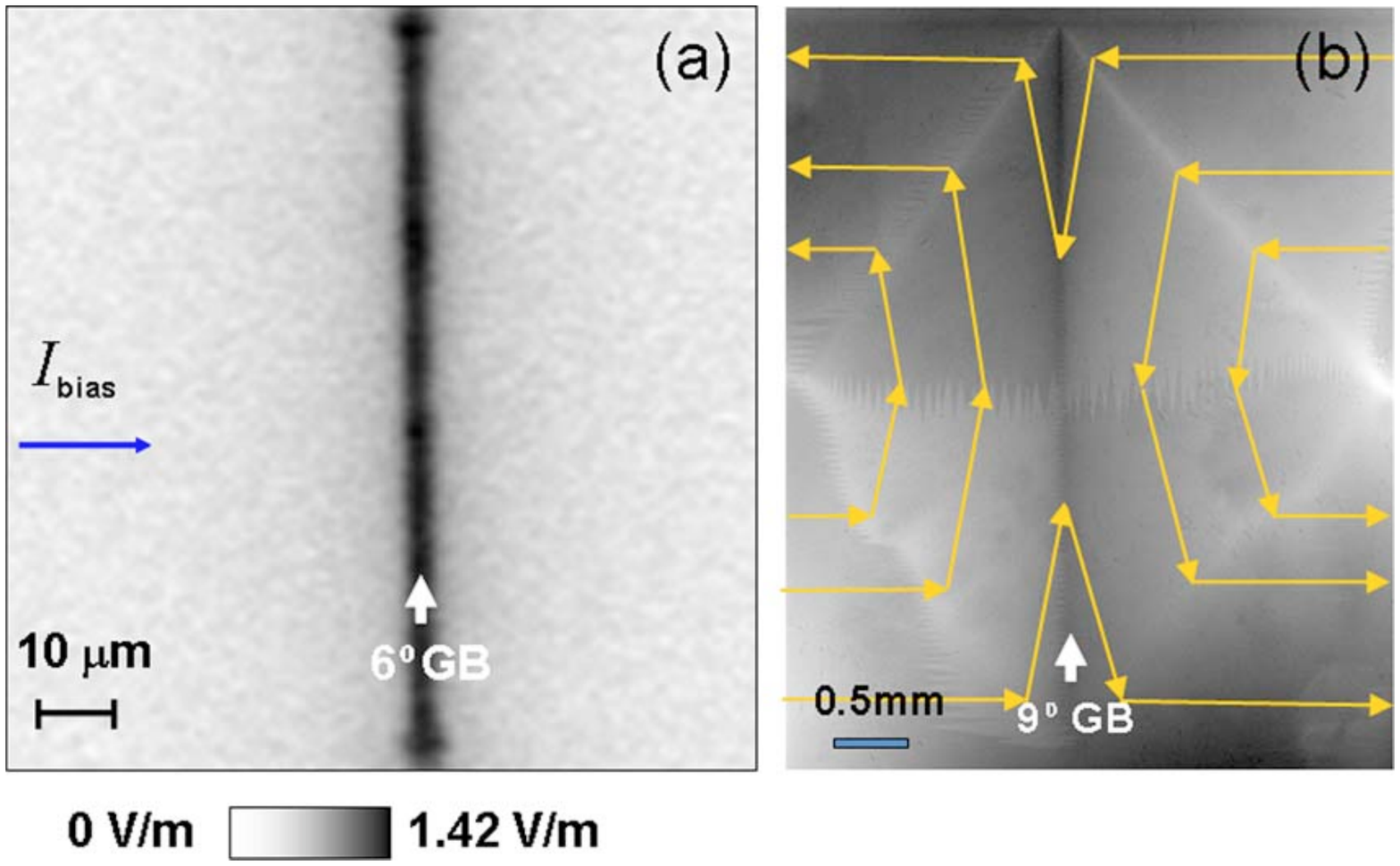}
		\caption{(a) LTLSM image scanned in 1\,$\mu$m steps of a 6$^\circ$ [001]-tilt bicrystal taken at 12\,K and 0.25\,T with bias current of 39\,mA which generated 17\,V
		across the bridge. (b) MO image of a 9$^\circ$ bicrystal with computed streamlines yellow of the magnetization current calculated from the Bean model.
		The MO image was taken after cooling to 6\,K in a perpendicular field of 120\,mT and then reducing the field to zero. Reproduced from $Appl$. $Phys$. $Lett$. {\bf95}, 212505 (2009), with the permission from AIP Publishing.} 
\label{fig:figure8}
\end{figure}

Already in the early stage of thin film research of pnictide superconductors, S. Lee $et$ $al.$ reported on the weak-link behavior of [001] tilt GBs in Co-doped Ba-122.\cite{Lee-3} They fabricated 350-nm thick films ($T_{\rm c}$ over 20\,K) on SrTiO$_3$ bicrystals ($\theta_{\rm GB}=3^\circ, 6^\circ, 9^\circ$ and 24$^\circ$) by PLD. Low-temperature laser scanning microscopy (LTLSM) and magneto-optical imaging (MOI) on these Co-doped Ba-122 bicrystal junctions with $\theta_{\rm GB}=6^\circ$ and 9$^\circ$ revealed the weak-link behavior also for low-angle GBs (fig.\,\ref{fig:figure8}). As can be seen in fig.\,\ref{fig:figure8}(a), a large dissipation across the 6$^\circ$ grain boundary was detected by LTLSM. For the 9$^\circ$ grain boundary sample, two clear roof-top shapes separated by the grain boundary are observed by MOI (fig.\,\ref{fig:figure8}(b)). These results indicated that the critical angle $\theta_{\rm c}$ may be below 6$^\circ$, which differs from results reported two years later by Katase $et$ $al$.\cite{Katase-1} We speculate that oxygen impurities may have limited the $J_{\rm c}^{\rm GB}$ for these Co-doped Ba-122 films on SrTiO$_3$. In fact, it is likely that the nanorods (see subsection {\it BaFeO$_3$ nanopillars}) tend to segregate along the GB reducing the effective cross-section and consequently $J_{\rm c}$. Detailed microstructural investigations are necessary to identify such differences.

\begin{figure}[ht]
	\centering
			\includegraphics[width=12cm]{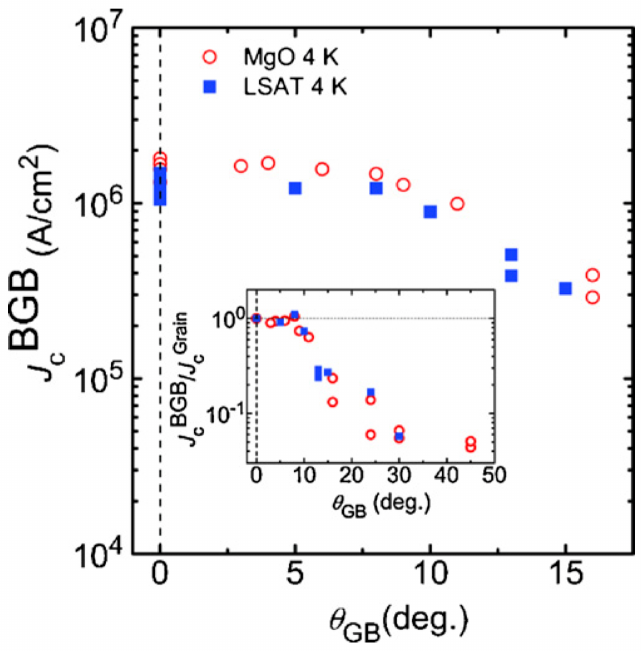}
		\caption{The $\theta_{\rm GB}$ dependence of inter-grain $J_{\rm c}$ for Co-doped Ba-122 bicrystals measured at 4\,K without magnetic field. Reproduced with permission from $Mater$. $Sci$. $Eng$., $B$ {\bf 177}, 515-519 (2012). Copyright 2012 Elsevier.} 
\label{fig:figure9}
\end{figure}

Later Katase,  Hiramatsu $et$ $al.$ reported a $\theta_{\rm c}$ as large as 9$^\circ$ from their detailed bicrystal experiments on Co-doped Ba-122.\cite{Katase-1, Hiramatsu-2} 
Here the respective $T_{\rm c}$ of Co-doped Ba-122 were 20.7\,K on MgO and 21.6\,K on LSAT bicrystals. Shown in fig.\,\ref{fig:figure9} is the $\theta_{\rm GB}$ dependence of inter-grain $J_{\rm c}$ measured at 4\,K without magnetic field. Clearly, the inter-grain $J_{\rm c}$ remained at 1\,MA/cm$^2$ up to $\theta_{\rm GB}=9^\circ$. Beyond this angle, the decay of inter-grain $J_{\rm c}$ sets in. Another interesting feature is that $J_{\rm c}^{\rm GB}$ was almost constant for $\theta_{\rm GB}=30^\circ$ and $\theta_{\rm GB}=45^\circ$.

\begin{figure}[ht]
	\centering
			\includegraphics[width=12cm]{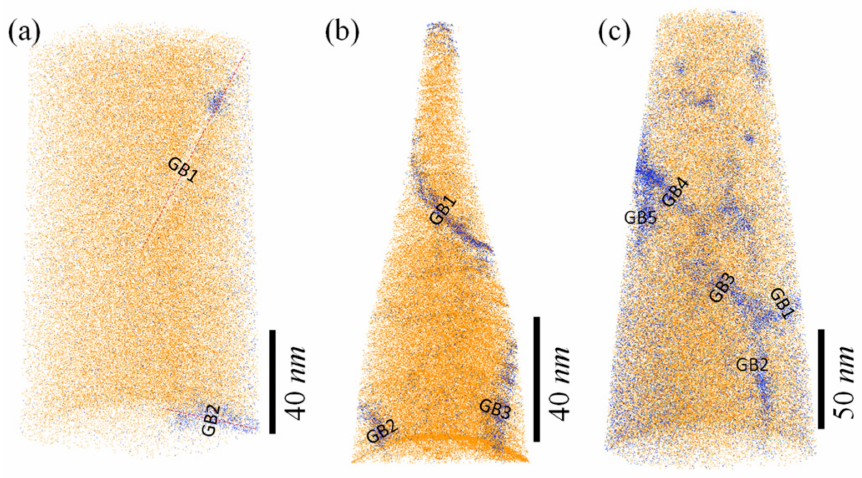}
		\caption{A 3-D atom-probe tomographic reconstruction of: (a) (Ba$_{0.6}$K$_{0.4}$)Fe$_2$As$_2$; (b) (Ba$_{0.6}$K$_{0.4}$)Fe$_2$As$_2$; and (c) Ba(Fe$_{0.92}$Co$_{0.08}$)$_2$As$_2$ superconductors. Oxygen atoms are in blue and Ba atoms are in orange, other elements are excluded for a clear display of grain boundary segregation. Each dot represents a single atom, but not to scale. Reproduced from $Appl$. $Phys$. $Lett$. {\bf105}, 162604 (2014), with the permission from AIP Publishing.} 
\label{fig:figure10}
\end{figure}

A high inter-grain $J_{\rm c}$ over 1\,MA/cm$^2$ even at $\theta_{\rm GB}=24^\circ$ and 4\,K was reported by Sakagami $et$ $al$.\cite{Sakagami} for a P-doped Ba-122 thin film grown by molecular beam epitaxy (MBE) on a MgO bicyrstal substrate. Although the number of data points is not sufficient, P-doped Ba-122 most likely also has a higher critical angle than the cuprates.

It was reported that polycrystalline K-doped Ba-122 bulk samples showed a large self-field $J_{\rm c}$ of $\sim0.1\,{\rm MA/cm}^2$.\cite{Weiss} Microstructural analysis revealed that the grain boundaries were clean without segregation of impurities. Later, atom probe tomography performed on both K- and Co-doped Ba-122 polycrystals revealed that traces of oxygen impurities were present at the GBs, in particular for the Co-doped sample (fig.\,\ref{fig:figure10}).\cite{Kim} Such oxidized regions may hide the intrinsic characteristics of the grain boundary. Although the microstructures of Co-doped and K-doped Ba-122 polycrystalline samples are almost identical, the in-field $J_{\rm c}$ performance of Co-doped Ba-122 is inferior to that of K-doped Ba-122 at 4.2\,K.\cite{Weiss} These results indicate that grain boundary properties depend on the kind of doping elements (i.e., K, Co and P) and perhaps $AE$ in $AE$Ba$_2$As$_2$ too. Therefore, grain boundary experiments on K-doped Ba- and Sr-122 are highly desirable and also important for on-going development of powder-in-tube processed wires and bulk magnets.\cite{Weiss-2}

\subsubsection{NdFeAs(O,F)}
\begin{figure}[!hb]
	\centering
			\includegraphics[width=12cm]{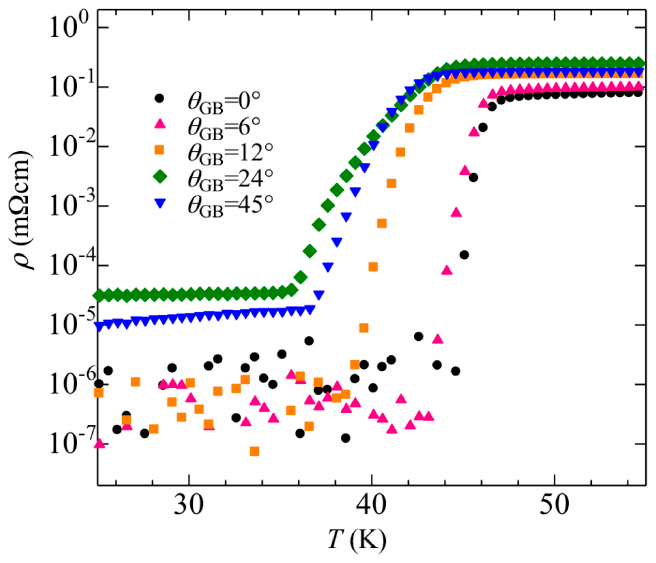}
		\caption{Semi-logarithmic plots of the resistivity curves for NdFeAs(O,F) bicrystals having different $\theta_{\rm GB}$. The $\theta_{\rm GB}=0^\circ$ represents the intra-grain $J_{\rm c}$. The figure is redrawn with some modifications based on the data taken from fig.\,3(b) in Ref.\,\onlinecite{Omura}.} 
\label{fig:figure11}
\end{figure}

$Ln$FeAs(O,F) ($Ln={\rm Sm}$ and Nd) compounds have the highest $T_{\rm c}$ among the FBS. Hence, a lot of effort was devoted to producing $Ln$FeAs(O,F) wires. However, the self-field $J_{\rm c}$ of polycrystalline SmFeAs(O,F) was only $\sim4\times10^4\,{\rm A/cm}^2$ at 4.2\,K,\cite{Zhang} which is one order of magnitude lower than for K-doped Ba-122, indicative of possible weak-link behavior. Hence bicrystal experiments on $Ln$FeAs(O,F) have been long desired to elucidate this question.

Omura $et$ $al.$ have recently reported on the GB angle dependence of transport properties for NdFeAs(O,F) thin films fabricated by MBE.\cite{Omura} For $\theta_{\rm GB}=6^\circ$ and 12$^\circ$, the resistivity in the superconducting state was below the instrumental limitation(fig.\,\ref{fig:figure11}). For $\theta_{\rm GB}\geq24^\circ$, in contrast a finite resistivity below $T_{\rm c}$ was observed and hence the inter-grain $J_{\rm c}$ was zero. The inter-grain $J_{\rm c}$ for $\theta_{\rm GB}$=6$^\circ$ was reduced by nearly 30\% compared to the intra-grain $J_{\rm c}$. Hence the critical angle for these NdFeAs(O,F) films was less than 6$^\circ$. However, the authors stated that the results did not reflect the intrinsic grain boundary properties due to the erosion of the grain boundaries by fluorine diffusion.

\subsubsection{Summary of the misorientation dependence of inter-grain $J_{\rm c}$}
\begin{figure}[ht]
	\centering
			\includegraphics[width=13cm]{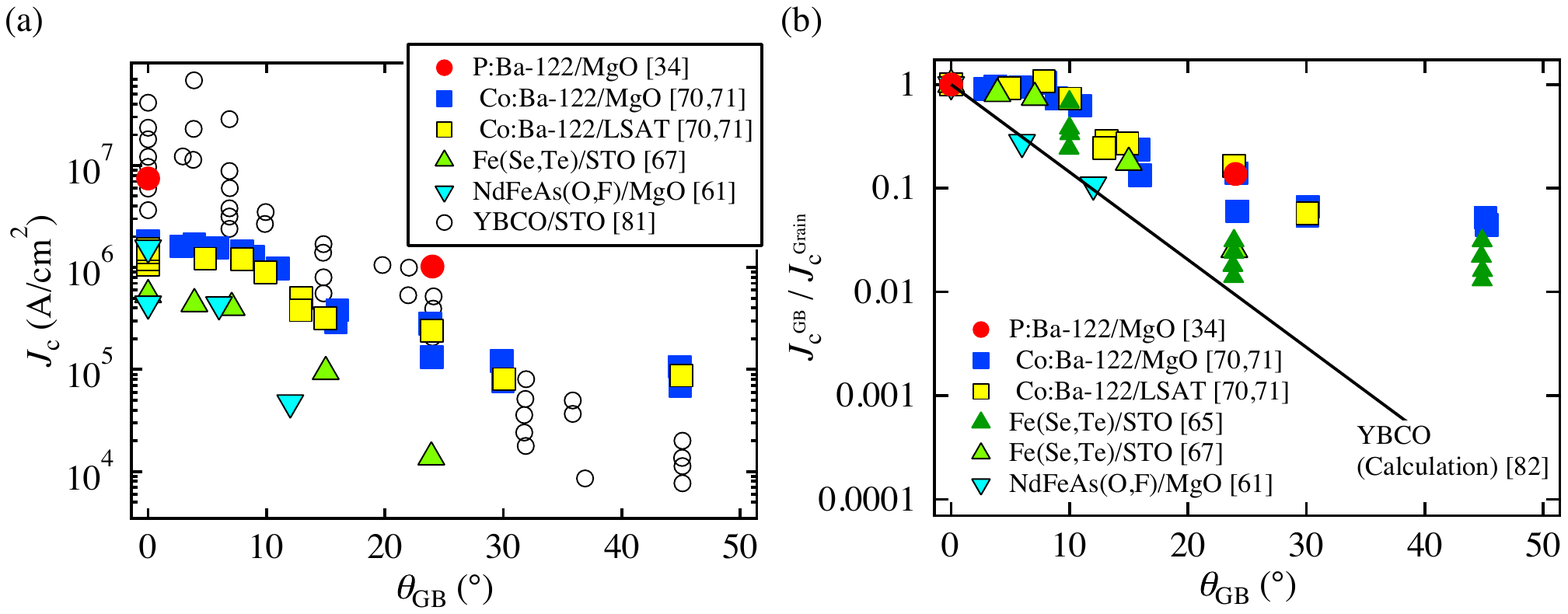}
		\caption{The $\theta_{\rm GB}$ dependence of inter-grain $J_{\rm c}$ for various FBS bicrystals. The data were acquired at 4\,K for P:Ba-122\,\cite{Sakagami} and Co:Ba-122\,\cite{Katase-1, Hiramatsu-2}, whereas for Fe(Se,Te)\,\cite{Sarnelli-2, Si-1} and NdFeAs(O,F)\,\cite{Omura} the measurements were conducted at 4.2\,K. For comparison, the data of YBa$_2$Cu$_3$O$_7$ (YBCO) measured at 4\,K were also plotted.\cite{Hilgenkamp}
		(b) The ratio of $J_{\rm c}^{\rm GB}/J_{\rm c}^{\rm Grain}$ as a function of $\theta_{\rm GB}$ for various FBS. The black line was YBCO taken from Ref.\,\onlinecite{Graser}.} 
\label{fig:figure12}
\end{figure}

Shown in fig.\,\ref{fig:figure12} is a summary of $\theta_{\rm GB}$ dependence of inter-grain $J_{\rm c}$ for various FBS bicrystals. Clearly, P-doped Ba-122 showed a high inter-grain $J_{\rm c}$ of 1\,MA/cm$^2$ at $\theta_{\rm GB}=24^\circ$ (fig.\,\ref{fig:figure12}(a)), which is highly beneficial for wire applications. It is also noted that $J_{\rm c}^{\rm GB}$ for Ba-122 is higher than those for YBCO beyond $\theta_{\rm GB}=24^\circ$.
Although intra-grain and inter-grain $J_{\rm c}$ are high for P-doped Ba-122, the $\theta_{\rm GB}$ dependence of $J_{\rm c}^{\rm GB}$ is similar to that of Co-doped Ba-122 (fig.\,\ref{fig:figure12}(b)). For all FBS except for NdFeAs(O,F), the critical angle $\theta_{\rm c}$ is around 9$^\circ$, which indicates that in-plane misorientation alignments of FBS grains less than 9$^\circ$ usually do not impede the critical current flow. Hence one can expect that
grain boundary networks whose in-plane misorientation is less than 9$^\circ$ work as flux pinning centers. Indeed, the pinning force density analysis of P-doped Ba-122 on technical substrates revealed that the GB pinning contribution is dominant for $H\parallel c$,\cite{Sato-2, Iida-3} which will be discussed later.

The initial slope for the $J_{\rm c}$ decrease with angle for FBS is almost similar to that of YBCO, fig.\,\ref{fig:figure12}(b), which is different from the common belief.
A distinct difference between FBS and YBCO is recognized above $\theta_{\rm GB}=24^\circ$: for YBCO $J_{\rm c}^{\rm GB}$ continuously decreases exponentially with $\theta_{\rm GB}$, whilst $J_{\rm c}^{\rm GB}$ remains constant for FBS. This may be due to the different symmetry: $d$-wave for YBCO and sign reversal $s$-wave for FBS. Additionally, the grain boundaries for both Ba-122 and Fe(Se,Te) are of metallic nature. Another plausible reason may be 45$^\circ$ [001]-tilt GBs in Ba-122 and Fe(Se,Te) compounds being special,
low-$\Sigma$ or facetted GBs with a narrower structurally and electronically disturbed region than for GBs with smaller $\theta_{\rm GB}$ as discussed for high--$T_{\rm c}$ cuprates.\cite{Chan}

It is reported that inter-grain $J_{\rm c}$ for YBCO is ameliorated by Ca-doping for which the depletion of the carrier density at GBs is compensated.\cite{Hammerl}
Later, electron energy loss spectroscopy (EELS) revealed high concentrations of Ca at the tensile regions of the GBs.\cite{Song}
On the assumption that the same mechanism could work for overdoped Ba-122, Durrell $et$ $al$. fabricated Co-doped Ba-122 bicrystals with Co concentration from 8\% (optimal) to 10\% (overdoped).\cite{John} However, no improvement of inter-grain $J_{\rm c}$ was observed.
More of such experiments should be conducted to tackle the GB problem for FBS.

\section{FBS thin films on technical substrates}

\begin{table*}[ht]
\caption{FBS thin films on technical substrates reported by BNL (Brookhaven National Laboratory), IEE-CAS (Institute of Electrical Engineering, Chinese Academy of Sciences), NIMS (National Institute of Materials Science), IFW (Leibniz Institute for Solid State and Materials Research Dresden), TIT (Tokyo Institute of Technology), and Nagoya (Nagoya University). LMO and EC depo. stand for LaMnO$_3$ and Electrochemical deposition, respectively.}
\small
 \begin{ruledtabular}
  \begin{tabular}{cccccc}  
Group &  Materials & Substrates & Methods &  $T_{\rm c} (K)$ & $J_{\rm c}$ performance \\ \hline
BNL  & Fe(Se,Te)\cite{Si-2} & IBAD-MgO & PLD  &  $T_{\rm c,R=0}=11$ & \begin{tabular}{c}$>1\times10^4\,{\rm A/cm}^2$\\ at 25\,T, 4.2\,K \end{tabular}\\
          &   Fe(Se,Te)\cite{Si-3} & \begin{tabular}{c} RABiTS\\ with CeO$_2$  \end{tabular}  & PLD & $T_{\rm c}^{\rm onset}=20$ & \begin{tabular}{c}0.1\,${\rm MA/cm}^2$\\ at 30\,T, 4.2\,K\end{tabular} \\ \hline
IEE-CAS & Fe(Se,Te)\cite{Xu} & \begin{tabular}{c}IBAD-MgO\\ with LMO\end{tabular} & PLD &  16.8 &  \begin{tabular}{c}0.43\,${\rm MA/cm}^2$\\ at self-field, 4.2\,K\end{tabular} \\ \hline
CNR & Fe(Se,Te)\cite{Braccini} & \begin{tabular}{c}RABiTS\\ with CeO$_2$\end{tabular} & PLD &  18 &  \begin{tabular}{c}$>2\times10^4\,{\rm A/cm}^2$\\ at 18\,T, 4.2\,K \end{tabular} \\ \hline        
NIMS  & FeSe\cite{Demura} & RABiTS & EC depo.    &   8  & N/A \\
           & FeSe\cite{Yamashita} & RABiTS & EC depo. &   \begin{tabular}{c}$T_{\rm c}^{\rm onset}=8.4$\\ $T_{\rm c,R=0}=2.5$\end{tabular}  & N/A \\ \hline
IFW    & Co:Ba-122\cite{Iida-2} & IBAD-MgO & PLD & $T_{\rm c}^{\rm onset}=21$ &\begin{tabular}{c} 0.1\,${\rm MA/cm}^2$\\ at self-field, 8\,K\end{tabular} \\
           & Co:Ba-122\cite{Trommler} & IBAD-MgO & PLD & $T_{\rm c,50\%}=23$ &\begin{tabular}{c} 0.1\,${\rm MA/cm}^2$\\ at 9\,T, 4.2\,K\end{tabular} \\ \hline 
TIT     & Co:Ba-122\cite{Katase-2} & IBAD-MgO & PLD & $T_{\rm c}^{\rm onset}=22.0$ &\begin{tabular}{c} 0.1\,${\rm MA/cm}^2$\\ at 9\,T, 4.2\,K\end{tabular} \\ \hline
IEE-CAS & Co:Ba-122\cite{Zhongtang-Xu} & \begin{tabular}{c}IBAD-MgO\\ with LMO\end{tabular}  & PLD & $T_{\rm c,onset}=20.2$ &\begin{tabular}{c} 0.86\,${\rm MA/cm}^2$\\ at 9\,T, 4.2\,K\end{tabular} \\ \hline
TIT     & P:Ba-122\cite{Sato-2} & IBAD-MgO  & PLD & $T_{\rm c}^{\rm onset}=26$ &\begin{tabular}{c}  $>0.11\times10^4\,{\rm A/cm}^2$\\ at 9\,T, 4.2\,K  \end{tabular} \\ 
          & P:Ba-122\cite{Sato-2} & IBAD-MgO  & PLD & $T_{\rm c}^{\rm onset}=23$ &\begin{tabular}{c} $0.2\times10^4\,{\rm A/cm}^2$\\ at 9\,T, 4.2\,K\end{tabular}  \\
TIT     & P:Ba-122\cite{Hosono-1} &\begin{tabular}{c} IBAD-MgO\\ 15\,cm long\end{tabular}  & PLD & $T_{\rm c,10\%}=18.0$ &\begin{tabular}{c} $I_{\rm c}=0.47$\,mA\\ at self-field \end{tabular} \\
          & P:Ba-122\cite{Hosono-1} &\begin{tabular}{c} IBAD-MgO\\ short tape \end{tabular}  & PLD & $T_{\rm c,10\%}=24.0$ &\begin{tabular}{c} $I_{\rm c}=975$\,mA\\ at self-field \end{tabular} \\
TIT     & P:Ba-122\cite{Iida-3} & IBAD-MgO  & PLD & $T_{\rm c,90\%}=28$ &\begin{tabular}{c} 0.1\,${\rm MA/cm}^2$\\ at 15\,T, 4.2\,K\end{tabular} \\ \hline              
Nagoya & NdFeAs(O,F)\cite{Iida-4} & IBAD-MgO & MBE & 43 &\begin{tabular}{c} $7\times 10^{4}\,{\rm cm}^{-2}$\\ at self-field, 5\,K\end{tabular} \\  
\end{tabular}
\end{ruledtabular}
\label{tab:FBS_tape}
\end{table*}

As described in the previous section, grain boundaries are less critical an issue for FBS than for cuprates, and powder-in-tube-processed wires based on {\it 11}, {\it 122} and {\it 1111} have been fabricated. Among them, the in-field $J_{\rm c}$ of {\it 122} wires has already reached the practical level of $10^5\,{\rm A/cm}^2$,\cite{Huang-He} and a 120\,m long tape of K-doped Sr-122 has been realized.\cite{Zhang-x} However, biaxially textured films on single-crystalline substrates and also single crystals\,\cite{Ishida} showed superior $J_{\rm c}\mathchar`-H$ performance to polycrystalline samples. To date, seven groups have published results on FBS films on technical substrates as prototype coated conductor samples, as summarized in Table\,\ref{tab:FBS_tape}. 
In this section, various FBS thin films on technical substrates are reviewed.  Most of these studies were done on IBAD-MgO templates (textured MgO templates are prepared on Hastelloy by ion beam assisted deposition), supplied by iBeam inc., USA and Shanghai Creative Supercond. Technol. Co. Ltd, China. Three studies on FeSe and Fe(Se,Te) used RABiTS (rolling-assisted biaxially textured substrate) template for which two of them were supplied by ENEA (Italian National Agency for New Technologies, Energy and Sustainable Economic Development) and evico GmbH, Germany. The other has been fabricated by in-house R\&D collaboration between Brookhaven National Laboratory and Oak Ridge National Laboratory.

\subsection{FeSe and Fe(Se,Te)}
Despite the large lattice mismatch between Fe(Se,Te) ($a_{\rm Fe(Se,Te)}=3.81$\,\AA) and MgO ($a_{\rm MgO}=4.21$\,\AA), epitaxial Fe(Se,Te) thin films were successfully
fabricated on hastelloy tapes, on which epitaxial MgO templates were deposited by IBAD.\cite{Si-2} The respective in-plane and out-of-plane full width at half maximum of Fe(Se,Te) were ${\Delta \phi=4.5^\circ}$ and ${\Delta \omega=3.5^\circ}$. Clearly, ${\Delta \phi=4.5^\circ}$ is less than the critical angle $\theta_{\rm c}$ for [001]-tilt Fe(Se,Te) grain boundaries.
Although a critical angle for [010]-tilt Fe(Se,Te) grain boundary junctions has not been reported, it is expected that [001]-tilt and [010]-tilt GBs in Fe(Se,Te) have a similar critical angle due to the small anisotropy and likely the sign reversal $s$-wave type symmetry. Therefore, weak-link behavior is absent in these Fe(Se,Te) films on IBAD-MgO. The lower $T_{\rm c}$ ($T_{\rm c,R=0}=11$\,K) of that film compared to the film on LaAlO$_3$ substrate ($\sim15$\,K) is mostly ascribed to the large lattice misfit. The Fe(Se,Te) film showed an almost isotropic $J_{\rm c}$ over 10$^4$\,A/cm$^2$ at 25\,T and 4.2\,K.
The pinning force analysis revealed that point defect pinning governs the $J_{\rm c}\mathchar`-H$ properties.

\begin{figure}[ht]
	\centering
			\includegraphics[width=7cm]{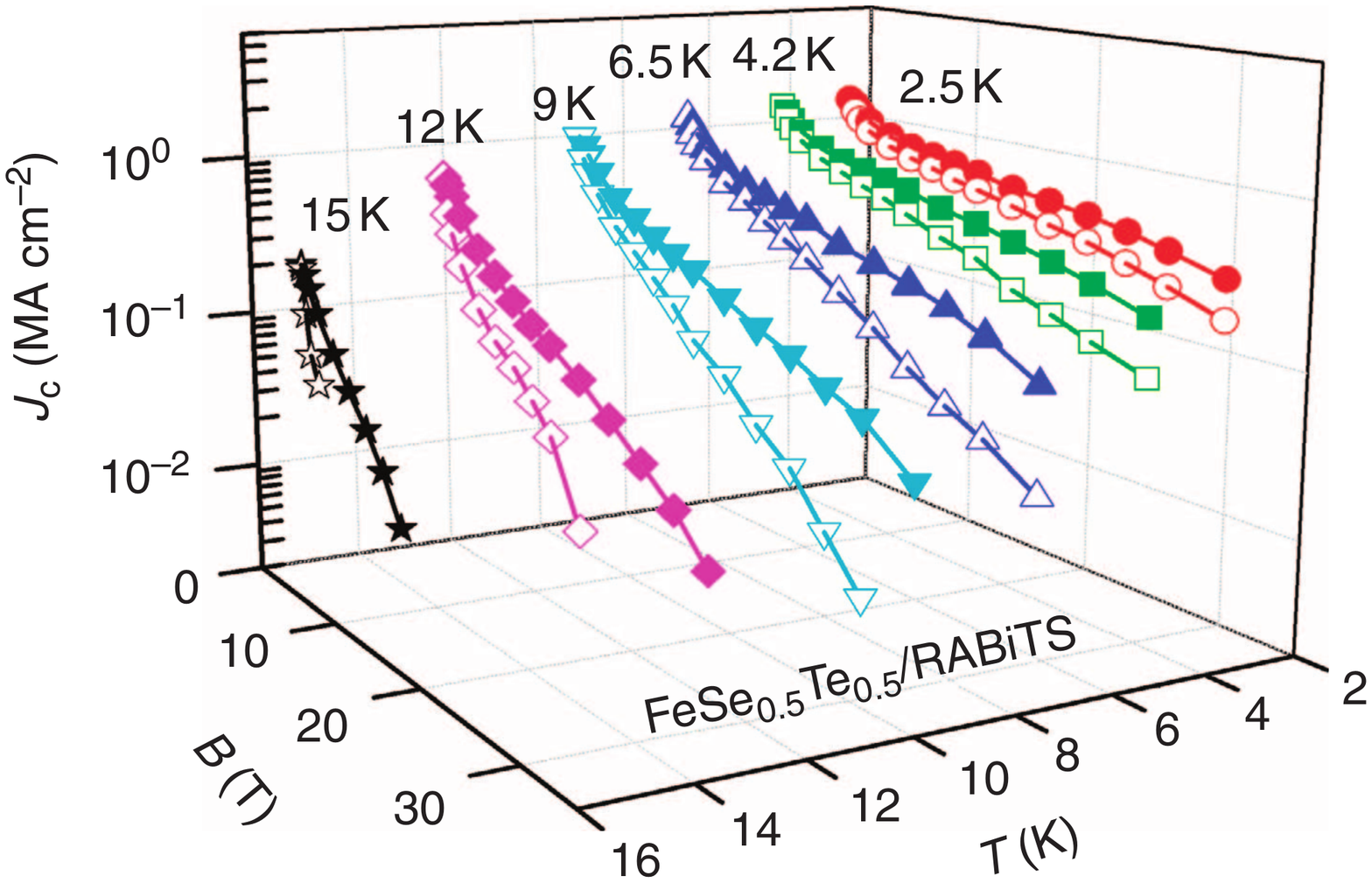}
		\caption{The field dependence of $J_{\rm c}$ for an Fe(Se,Te) film deposited on RABiTS measured at various temperatures.
		Closed and open symbols represent $H\parallel c$ and $\parallel ab$, respectively. Reproduced with permission from $Nat$. $Commun$. {\bf 4}, 1347 (2013). Copyright 2013 Spring Nature.} 
\label{fig:figure13}
\end{figure}

To further improve $J_{\rm c}\mathchar`-H$ properties of Fe(Se,Te) on technical substrate, Si $et$ $al$. have fabricated Fe(Se,Te) thin films on CeO$_2$-buffered RABiTS.\cite{Si-3} The lattice parameter of CeO$_2$ is around 3.82\,\AA ($\sim5.41/\sqrt{2}$), which almost perfectly matches the in-plane lattice parameter of Fe(Se,Te). As a result, Fe(Se,Te) on RABiTS showed a high onset $T_{\rm c}$ of over 20\,K with a sharp transition of 1\,K width. The structural characterization by X-ray diffraction revealed an in-plane FWHM ${\Delta \phi_{\rm Fe(Se,Te)}=6^\circ}$. Despite the larger in-plane texture spread, the Fe(Se,Te) film was capable of carrying large current densities (fig.\,\ref{fig:figure13}). A nearly isotropic
$J_{\rm c}\sim0.1$\,MA/cm$^2$ at 4.2\,K and 30\,T was observed. The pinning mechanism was the same as for the Fe(Se,Te) film on IBAD-MgO. The authors concluded that the use of
CeO$_2$ buffer layer is the key for achieving high $T_{\rm c}$. Additionally, the substrates containing low angle grain boundaries up to 7$^\circ$ do not impede the critical current flow, which is different from $RE$Ba$_2$Cu$_3$O$_7$.

Quite recently, the group of IEE-CAS reported Fe(Se,Te) thin films on IBAD-MgO with LaMnO$_3$ (LMO) buffer layers.\cite{Xu} The Fe(Se,Te) was grown on LMO with cube-on-cube configuration. The respective out-of-plane and in-plane full width at half maximum values of Fe(Se,Te) are $\Delta \omega=3.4^\circ$ and $\Delta \phi=7.8^\circ$. The $T_{\rm c}$ of the sample is 16.8\,K, which is slightly higher than that of bulk due to the compressive strain. Albeit the large $\Delta \phi$ of 7.8$^\circ$, an in-field $J_{\rm c}$ above 0.35\,${\rm MA/cm}^2$ up to 9 T was recorded at 4.2\,K. Additionally, the anisotropy of $J_{\rm c}$ was relatively small.

The group of CNR has been recently working on an attempt to decrease the number of necessary buffer layers or exchanging the metallic substrate to less expensive materials. Both solutions could reduce the fabrication costs of such Fe(Se,Te) coated conductors.\cite{Braccini} The films on Ni5W RABiTS covered with a single CeO$_2$ buffer layer (produced by ENEA) showed a $T_{\rm c}$ of $\sim$18\,K and isotropic $J_{\rm c}$ above $2\times10^4$\,A/cm$^2$ at 4.2\,K in fields up to 18\,T. On biaxially textured invar (nickel-iron alloy with 36 mass\% of Ni) tape, epitaxial Fe(Se,Te) layers of the same texture quality as the tape itself (FWHM$\sim$ 9$^\circ$) were grown without any buffer layer. The superconducting properties of these samples still have to be investigated.

The group at NIMS proposed a unique technique for fabricating FeSe films on RABiTS provided by evico GmbH.\cite{Demura, Yamashita} They performed electrochemical deposition using three electrodes, Pt (counter electrode), Ag/AgCl (reference electrode) and RABiTS (working electrode) at 70$^\circ$C. The electrolyte was prepared from a solution of FeCl$_2\cdot4$H$_2$O, SeO$_2$, and Na$_2$SO$_4$ in distilled water. A pH value of 2.1 for the electrode was maintained during the deposition. X-ray diffraction confirmed FeSe phase formation without heat treatment. Although the resulting film had a low $T_{\rm c}$ of 2.5\,K, this technique is promising for further development and optimization.\cite{Yamashita}

\subsection{Co- and P-doped Ba-122}
The first report on Co-doped Ba-122 on technical substrates was published by Iida $et$ $al$.\cite{Iida-2} They fabricated Co-doped Ba-122 films on Fe-buffered IBAD-MgO templates by PLD. Due to the relatively large in-plane mosaicity of the MgO (${\Delta \phi_{\rm MgO}\sim6^\circ}$), the crystalline quality of the intermediate Fe as well as Co-doped Ba-122 layers was poor (i.e., ${\Delta \phi_{\rm Fe}\sim5^\circ}$ and ${\Delta \phi_{\rm Ba-122}\sim5^\circ}$). Although biaxially textured Co-doped Ba-122 showed an onset $T_{\rm c}$ of 21\,K, the in-field $J_{\rm c}$ at 8\,K was almost one order of magnitude lower compared to the film on single crystalline MgO due to the poor crystalline quality. Later, employing a high-quality MgO template (${\Delta \phi_{\rm MgO}\sim2.4^\circ}$) improved the in-field $J_{\rm c}$ of Co-doped Ba-122 to a level almost comparable to films on single crystalline substrates.\cite{Trommler}

Almost at the same time, Katase $et$ $al.$ also reported on the fabrication of Co-doped Ba-122 directly on IBAD-MgO.\cite{Katase-2} A sharp texture of the Co-doped Ba-122
with ${\Delta \phi_{\rm Ba122}\sim3.2^\circ-3.5^\circ}$ was obtained irrespective of the in-plane mosaicity of MgO templates due to the self-epitaxy effect. The onset $T_{\rm c}$ of Co-doped Ba-122 on IBAD-MgO was 22\,K. In-field $J_{\rm c}$ reached almost 0.1\,MA/cm$^2$ under 9\,T at 4\,K.

Very recently, Co-doped Ba-122 was also fabricated on IBAD-MgO with LMO by Xu $et$ $al$.\cite{Zhongtang-Xu} They used SrTiO$_3$ as buffer layer between LMO and Co-doped Ba-122. The resultant film showed almost isotropic $J_{\rm c}$. The respective $J_{\rm c}$ for $H\parallel c$ and $\parallel ab$ were 0.86 and 0.96\,MA/cm$^2$ at 9\,T and 4.2\,K.
 
\begin{figure}[ht]
	\centering
			\includegraphics[width=10cm]{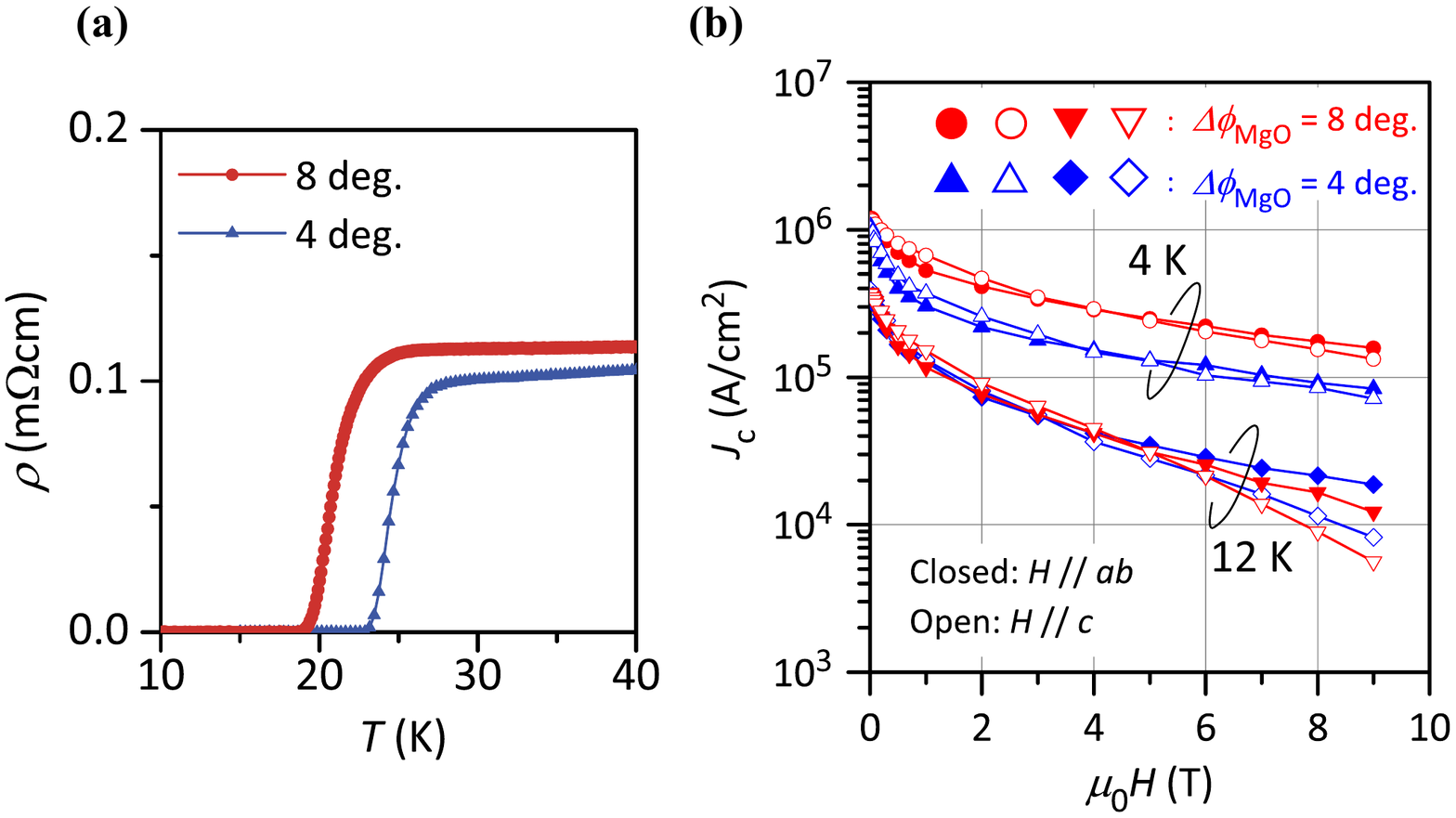}
		\caption{(a) The temperature dependence of resistivity for P-doped Ba-122 deposited on IBAD-MgO with different  in-plain mosaicities, ${\Delta \phi_{\rm MgO}=4^\circ}$ and 8$^\circ$. (b) The corresponding in-field $J_{\rm c}\mathchar`-H$ properties of these films measured at 4\,K and 12\,K. Reproduced from Sato $et$ $al$., $Sci$. $Rep$. {\bf 6}, 36828 (2016).} 
\label{fig:figure14}
\end{figure}

\begin{figure}[ht]
	\centering
			\includegraphics[width=10cm]{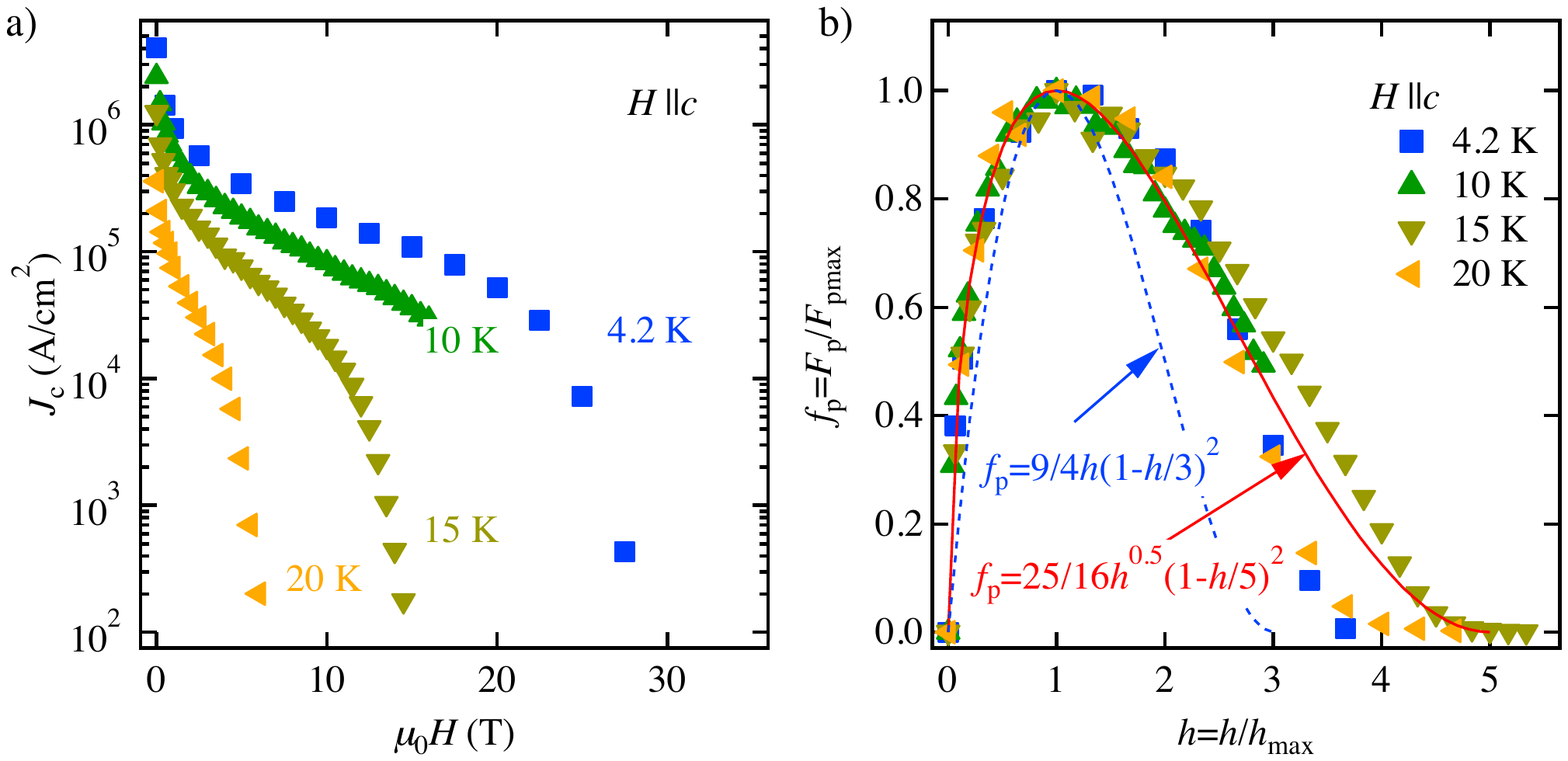}
		\caption{(a) $J_{\rm c}\mathchar`-H$ properties of a P-doped Ba-122 on IBAD-MgO measured at different temperatures for $H\parallel c$. (b) The corresponding normalized pinning force density ($f_{\rm p}$) as a function of reduced field $h$. The solid line is for 2D defects such as grain boundary, and the dotted line is for 3D defects such as point-like defect. The figure is redrawn with some modifications based on the data taken from fig.\,5 in Ref.\,\onlinecite{Iida-3}.} 
\label{fig:figure15}
\end{figure}

P-doped Ba-122 shows the second highest $T_{\rm c}$ among the 122 compounds and, therefore, $J_{\rm c}\mathchar`-H$ properties are usually superior to Co-doped Ba-122. Additionally, growth-related and artificial pinning centers can be introduced into the P-doped Ba-122 matrix as discussed in the previous section.
Sato $et$ $al.$ fabricated P-doped Ba-122 thin films with PLD on IBAD-MgO with two different in-plain texture qualities, ${\Delta \phi_{\rm MgO}=4^\circ}$ and 8$^\circ$.\cite{Sato-2} The structural characterization by X-ray diffraction revealed ${\Delta \phi_{\rm Ba-122}=2.7^\circ}$ and 8$^\circ$ for P-doped Ba-122 deposited on the well and poorly textured MgO template, respectively. The film on poor crystalline MgO template had an onset $T_{\rm c}$ of 23\,K, which is comparable to the Co-doped Ba-122 on IBAD-MgO
 (fig.\,\ref{fig:figure14}(a)).
On the other hand, the P-doped Ba-122 on the well textured MgO template had a higher $T_{\rm c}$ of 26\,K (fig.\,\ref{fig:figure14}(a)). 
The low-$T_{\rm c}$ sample showed nonetheless better in-field $J_{\rm c}$ properties: almost isotropic $J_{\rm c}$ exceeding 0.1\,MA/cm$^2$ at 9\,T and 4\,K (fig.\,\ref{fig:figure14}(b)). This can be explained by grain boundary pinning. Figure\,\ref{fig:figure15} shows the $J_{\rm c}\mathchar`-H$ properties of a further optimized P-doped Ba-122 thin film on IBAD-MgO ($T_{\rm c,90\%Rn}$ of 28.3\,K) and the corresponding normalized pinning force density ($f_{\rm p}$) as a function of $h=H/H_{\rm max}$, where $H_{\rm max}$ is the field at which the pinning force density ($F_{\rm p}$) shows the maximum value.\cite{Iida-3} As can be seen, $f_{\rm p}$ curves for $H\parallel c$ at different temperatures almost fall onto a master curve described by

\begin{equation}
f_{\rm p} = \frac{25}{16}h^{0.5}\left(1-\frac{h}{5}\right)^2
\end{equation}
\noindent
This formula is analogous to
\begin{equation}
\left(\frac{H}{H_{\rm irr}}\right)^p\left(1-\frac{H}{H_{\rm irr}}\right)^q
\end{equation}
\noindent
($p=0.5, q=2$, and $H_{\rm irr}$ is the irreversibility field) found by Dew-Hughes for pinning by 2D defects such as grain boundaries.\cite{Hughes} Recently, Paturi $et$ $al$. found that the exponent $p=0.5$ for YBCO regardless of $q$ for a defect size in the order of coherence length.\cite{Paturi} On the other hand, $p$ increases to 1 with increasing defect size. This confirms that the pinning in P-doped Ba-122 on IBAD-MgO is dominated by the dislocations with nano-size.

Hosono's group even fabricated a 10\,cm long P-doped Ba-122 tape using a reel-to-reel system by PLD,\cite{Hosono-1} although the growth conditions were not fully optimized. The resultant film showed a slightly lower $T_{\rm c,10\%Rn}$ of 18.7\,K and a low $I_{\rm c}$ of 0.47\,mA per cm-width at 4.2\,K, corresponding to a $J_{\rm c}$ of $4.7\times10^4$\,A/cm$^2$. They also found that even short samples had a low $T_{\rm c}$. To improve the superconducting properties, they fabricated Fe$_3$P/P-doped Ba-122 bilayers on IBAD-MgO. As a result, the superconducting properties were improved to $T_{\rm c,10\%Rn}=24$\,K and $I_{\rm c}=975$\,mA per cm-width ($J_{\rm c}=1.75\times10^5$\,Acm$^2$) at 4.2\,K.

\subsection{NdFeAs(O,F)}

\begin{figure}[hb]
	\centering
			\includegraphics[width=5cm]{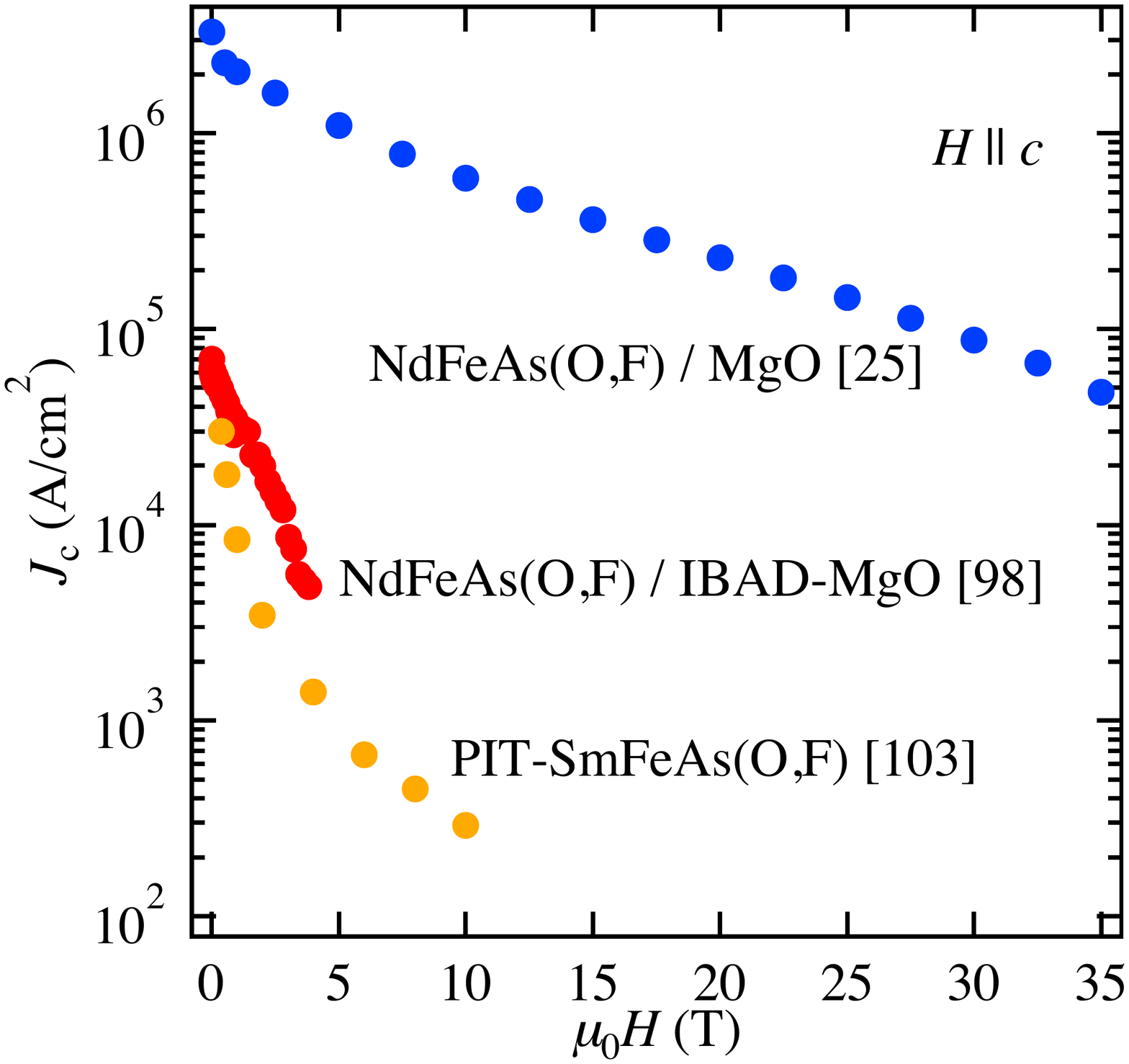}
		\caption{The $J_{\rm c}\mathchar`-H$ properties of the NdFeAs(O,F) on IBAD-MgO at 5\,K\cite{Iida-4} in comparison to the data of PIT-processed SmFeAs(O,F) wires\cite{Q-Zhang} and a NdFeAs(O,F) thin film on single crystal MgO\cite{Tarantini-6} at 4.2\,K.} 
\label{fig:figure16}
\end{figure}

Biaxially textured NdFeAs(O,F) films on IBAD-MgO have been fabricated by MBE at Nagoya University.\cite{Iida-4} The 102 peak $\phi$-scan revealed a small amount of
45$^\circ$ rotated grains. Shown in fig.\,\ref{fig:figure16} are $J_{\rm c}\mathchar`-H$ properties of NdFeAs(O,F) for $H\parallel c$ in comparison to data on PIT-processed SmFeAs(O,F) wires and a NdFeAs(O,F) film on single crystal MgO. Here, $J_{\rm c}$ of NdFeAs(O,F) on IBAD-MgO was calculated from the hysteresis loops using the Bean model. The NdFeAs(O,F) thin film on IBAD template had a self-field
$J_{\rm c}$ of $7\times10^4$\,A/cm$^2$ at 5\,K, whereas the corresponding value for PIT-processed SmFeAs(O,F) was around $4\times10^4$\,A/cm$^2$ at 4.2\,K.\cite{Q-Zhang} Now the performance gap between the tape and wire is very small. However, given that $J_{\rm c}$ of NdFeAs(O,F) on single crystalline MgO substrates is above 10$^4$\,A/cm$^2$ at 4.2\,K even at fields as high as at 35\,T and self-field values reach 3\,MA/cm$^2$,\cite{Tarantini-6} there is certainly room for further improvement of superconducting properties of NdFeAs(O,F) on IBAD-MgO. The reason for the low $J_{\rm c}$ values on IBAD templates is most likely the strong F diffusion in the GB regions, as revealed by bicrystal experiments, see above.

\subsection{Summary of FBS thin films on technical substrates}

\begin{figure}[ht]
	\centering
			\includegraphics[width=10cm]{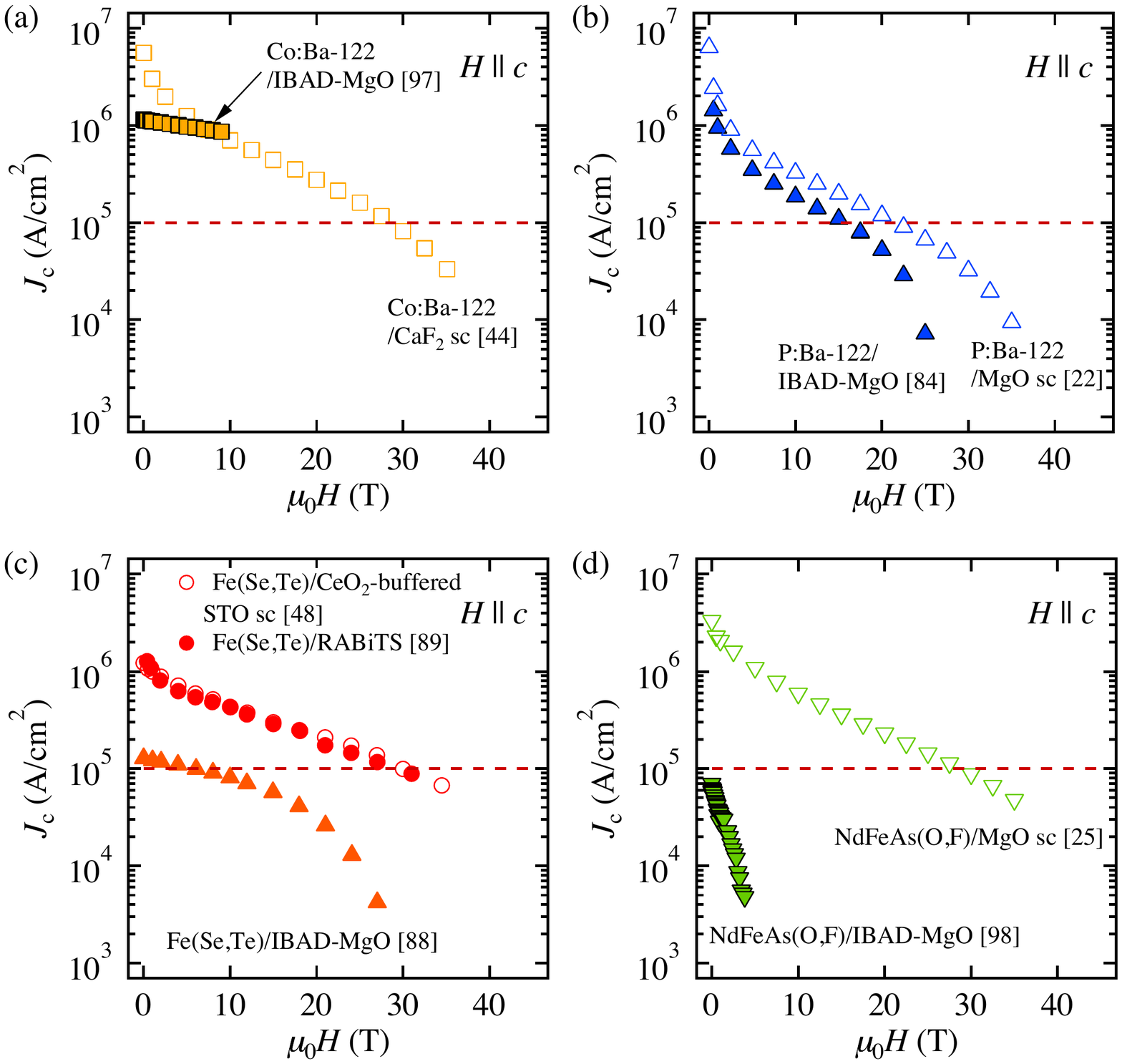}
		\caption{$J_{\rm c}\mathchar`-H$ properties of various FBS thin films on technical substrates and single crystalline substrates. Measurement temperature was 4.2\,K
		except for NdFeAs(O,F) on IBAD-MgO. The applied magnetic field was $H\parallel c$:
		(a) Co:Ba-122 on IBAD-MgO\,\cite{Zhongtang-Xu} and pinning enhanced Co:Ba-122 on CaF$_2$ fabricated by PLD.\cite{Tarantini-2}
		(b) P:Ba-122 on IBAD-MgO\,\cite{Iida-3} by PLD and MBE-processed P:Ba-122.\cite{Fritz}
		(c) Fe(Se,Te) on IBAD-MgO\,\cite{Si-2} and RABiTS.\cite{Si-3} Pinning enhanced Fe(Se,Te) on CeO$_2$-buffered SrTiO$_3$.\cite{Ozaki} All films were fabricated by PLD.
		(d) NdFeAs(O,F) on IBAD-MgO\,\cite{Iida-4} at 5\,K and MgO single crystalline substrate fabricated by MBE.\cite{Tarantini-6}}
\label{fig:figure17}
\end{figure}

Figure\,\ref{fig:figure17} compares $J_{\rm c}\mathchar`-H$ characteristics at 4.2\,K of various FBS thin films on technical substrates and of high-quality FBS thin films on single-crystalline substrates. For Co-doped Ba-122, although low-field $J_{\rm c}$ for the film on IBAD\,\cite{Zhongtang-Xu} is lower than that for the film on single crystal,\cite{Tarantini-2} a crossover is visible at 7.5\,T.
As can be seen in fig.\,\ref{fig:figure17}(b), in-field $J_{\rm c}$ performance for the P-doped Ba-122 on IBAD-MgO\cite{Iida-3} is slightly inferior to the film on MgO single-crystalline substrate\cite{Fritz}(fig.\,\ref{fig:figure17}(b)). Such a difference in $J_{\rm c}\mathchar`-H$ for the P-doped Ba-122 is mainly explained by the fact that the films on single-crystalline substrates are pinning-improved. It is expected that similar microstructures transferred to tapes would lead to similar $J_{\rm c}$ levels. Additionally, the phase formation has to be optimized so that the full potential of the films on single crystals can be achieved on tape. On the other hand, $J_{\rm c}\mathchar`-H$ of Fe(Se,Te) on RABiTS\,\cite{Si-3} is almost comparable to that of the pinning-enhanced Fe(Se,Te) thin film\,\cite{Ozaki} (fig.\,\ref{fig:figure17}(c)). Finally, the largest discrepancy is seen for NdFeAs(O,F) on IBAD-MgO and single crystalline MgO(fig.\,\ref{fig:figure17}(d)), which can be explained by the strong F diffusion along GBs and a concomitant broadening of their regions. This might be improved by employing flourine-free oxygen-deficient $Ln$FeAsO compounds.

To evaluate the best possible $J_{\rm c}$ performance, the ultimate critical current density, i.e., depairing current density at 4.2\,K ($J_{\rm d}(4.2\,{\rm K})$) for various FBS estimated by eqs.\,(4) and (5) is tabulated in table\,\ref{tab:physical}. 

\begin{equation}
J_{\rm d}(0) = \frac{\phi_0}{3\sqrt{3}\pi\mu_0\lambda_{\rm ab}^2(0)\xi_{\rm ab}(0)}
\end{equation}

\begin{equation}
J_{\rm d}(T) =J_{\rm d}(0)\left\{1-\left(\frac{T}{T_{\rm c}}\right)^2\right\}^{1.5}\left\{1+\left(\frac{T}{T_{\rm c}}\right)^2\right\}
\end{equation}

Clearly, Ba-122 compounds possess the highest $J_{\rm d}(4.2\,{\rm K})$ among the FBS. The maximum achievable $J_{\rm c}$ by core pinning is around 30\% of the depairing current density\,\cite{Paul-Seidel}. However, the pinning efficiency, here defined as $J_{\rm c}^{\rm s.f.}/J_{\rm d}$ at 4.2\,K for FBS even on single crystalline substrates reaches only $3.5\sim5.8\%$. 

\begin{table}[ht]
\caption{Physical quantities of Fe(Se,Te), Co- and P-doped Ba-122, and NdFeAs(O,F) used for the calculations of the
$J_{\rm c}^{\rm s.f.}/J_{\rm d}$ at 4.2\,K for FBS films on single crystals.}
 \small
 \begin{ruledtabular}
 \begin{tabular}{ccccc}
Materials                &  $\lambda_{\rm ab}(0)$ (nm)   & $\xi_{\rm ab}(0)$ (nm)  &  $J_{\rm d}$(4.2\,K) (MA/cm$^2$) & $J_{\rm c}^{\rm s.f.}/J_{\rm d}$ (\%) at 4.2\,K \\ \hline
Fe(Se,Te)               &  430\,\cite{Klein}                      & 1.5\,\cite{Klein}                    &  34     &    3.5            \\
Co-doped Ba-122  &  190\,\cite{Porozorov}             & 2.76\,\cite{Porozorov}      &  100   &    5.6     \\
P-doped Ba-122    &   200\,\cite{Chaparro}              & 2.14\,\cite{Chaparro}           &  117   &    5.4     \\
NdFeAs(O,F)         &   270\,\cite{Kacmarcik}            & 2.4\,\cite{Kacmarcik}            &  57     &    5.8      \\
\end{tabular}
\end{ruledtabular}
\label{tab:physical}
\end{table}

\section{Summary and discussion}
In order to understand the performance potential of FBS, we discussed the main application-related properties like pinning improvement by natural and artificial defects, the transparency of grain boundaries and the properties of GB networks. Naturally grown defects (BaFeO$_2$ in Co-doped Ba-122 and threading dislocations in Fe(Se,Te)) mainly contribute to the $c$-axis correlated pinning, and point pinning introduced by strain is beneficial as well. Artificial pinning centers can be introduced in the superconducting matrix by multilayer deposition, irradiation, and the addition of nanoparticles such as BaZrO$_3$ to the PLD targets. Although all methods described here are effective for improving in-field properties, the $J_{\rm c}$
level at 4.2\,K achieved so far is still only up to $\sim$6\% of the depairing current density in films on single crystals, suggesting room for further improvement of superconducting properties of FBS.

Bicrystal experiments on Ba-122 and Fe(Se,Te) revealed the metallic nature of grain boundaries. The critical angle above which inter-grain $J_{\rm c}$ falls off is around 9$^\circ$. However, the initial slope for the logarithmic $J_{\rm c}$ decrease with GB angle for FBS is similar to that of YBCO.
A distinct difference between FBS and YBCO is recognised above $\theta_{\rm GB}=24^\circ$: for YBCO $J_{\rm c}^{\rm GB}$ continuously decreases with $\theta_{\rm GB}$, whilst $J_{\rm c}^{\rm GB}$ remains constant for FBS. Overall, those properties are highly beneficial for making cheap FBS conductors.

The three main systems with largest application potential ({\it 11}, {\it 122} and {\it 1111}) have already been realized on technical substrates.
Most of the FBS coated conductor prototypes showed a comparable $T_{\rm c}$ to their counterparts on single-crystalline substrates. 
Fe(Se,Te) and P-doped Ba-122 reached the practical level (0.1\,MA/cm$^2$ at 15\,T). It is noted that PIT-processed FBS wires (K-doped Ba-122 and Sr-122) also have reached this level,\cite{Hosono-2} despite the presence of many GBs. Therefore, further GB experiments on FBS, especially for K-doped {\it122}, are desired. To further improve the in-field $J_{\rm c}$ properties of FBS on technical substrates, the introduction of artificial pinning centers without compromising $T_{\rm c}$ is necessary .

Even though FBS tapes will never compete with YBCO tapes regarding their superconducting properties, they could compete with YBCO in terms of cost. For YBCO tapes, expensive highly-textured metallic substrates and Ag layers to allow oxygenation are necessary. However, FBS can be realized on poorly textured templates without expensive cover layers. For realizing FBS conductors, there are other properties that need to be evaluated: i) thick FBS thin films ($\sim$1\,$\mu$m) on technical substrates should be fabricated and characterized for realizing FBS coated conductors. ii) Strain tolerance dependence of $J_{\rm c}$ should be investigated, which is important for making a magnet.

\begin{acknowledgments}
We thank for financial support from the EU (Iron-Sea under project number FP7-283141, SUPERIRON, Grant number 283204). This work was partially supported by the JSPS Grant-in-Aid for Scientific Research (B) Grant Number 16H04646. A portion of this work was performed at the National High Magnetic Field Laboratory, which was supported by National Science Foundation Cooperative Agreement No. DMR-1644779, and the State of Florida.
\end{acknowledgments}

\end{document}